\documentclass[pra,twocolumn]{revtex4}
\usepackage{amsmath}
\usepackage{amsfonts}
\usepackage{amssymb}
\usepackage{graphicx}
\usepackage{bm}
\begin{document}
\title{Normal modes for $N$ identical particles:
A study of the evolution of collective behavior from few-body to many-body}
\author{D. K. Watson \\
University of Oklahoma \\ Homer L. Dodge Department of Physics and Astronomy \\
Norman, OK 73019}
\date{\today}

\begin{abstract}
Normal mode dynamics are ubiquitous in nature underlying the motions of diverse
 interacting systems from the behavior of rotating stars to the vibration of 
crystal structures.  These behaviors are composed of simple collective motions 
of the $N$ interacting particles which move with the same frequency and phase,  
thus encapsulating many-body effects into simple dynamic motions.  In some 
regimes these collective motions are coupled by higher order effects and 
exhibit 
complex behavior, while in regimes such as the unitary regime for ultracold 
Fermi gases, a single collective mode can dominate, leading to quite simple 
behavior as seen in superfluidity.  In this study, I investigate the evolution
of collective motion as a function of $N$ for five types of normal modes 
obtained from an $L=0$ group
 theoretic solution of a general Hamiltonian for confined, identical particles.
  I show using simple analytic forms for the $N$-body normal modes 
that the collective behavior of few-body systems, which
 have the well known motions of molecular equivalents such as ammonia and 
methane, evolves smoothly as $N$ increases
to the collective motions expected for large $N$ ensembles (breathing, 
center of mass, particle-hole radial and angular excitations and phonon).  
Furthermore, the transition from few-body
 behavior (symmetric stretch, symmetric bend, antisymmetric stretch, 
antisymmetric bend and the opening and closing of alternative interparticle 
angles) to large $N$ behavior occurs at quite low values of $N$.  By $N=10$, 
the behavior of these normal modes has clearly become the expected large $N$
 behavior
 that describes ensembles with $10^{23}$   particles or more.  
I extend this investigation to a Hamiltonian known to 
support collective behavior, 
the Hamiltonian for Fermi gases in the unitary regime. I analyze 
both the evolution of the coefficients that mix the radial and angular 
symmetry coordinates in the expressions for the normal 
modes as well as the
 evolution of the normal mode frequencies as a function of $N$. Both quantities 
depend on interparticle interactions. This analysis reveals 
two phenomena that could contribute to the viability of collective behavior.  
First the coefficients that mix radial and angular coordinates in the normal 
modes go to zero or unity, i.e. no mixing, as $N$ becomes large resulting in 
solutions that do not depend on the details of the interparticle potential
as expected for this unitary regime, and that manifest the symmetry of an 
underlying approximate Hamiltonian.   
Second, the five normal mode frequencies which all start out 
close to the trap frequency for low values of $N$ ($N < 10$) separate rapidly 
as $N$ increases, creating large gaps between the normal modes that 
can, in principle, offer stability to collective behavior if mechanisms to 
prevent the transfer of energy to other modes exist (such as low temperature) 
or can be constructed. With the recent success using normal modes to
describe the emergence of 
collective behavior in the form of superfluidity in ultracold
Fermi gases in the unitary regime, understanding 
the character of these normal modes and
the evolution of their behavior as a function of $N$ has become of some
interest due to the possibility of offering
 insight into the dynamics  
of regimes supporting collective behavior. In this study, I investigate 
both the macroscopic behavior associated with these $N$-body normal modes, as
well as the microscopic motions underlying this behavior,
 and study the evolution of their collective behavior as a function of $N$.

\end{abstract}


\maketitle

\section{Introduction}

Normal modes occur in every part of the universe and at all scales from 
nuclear physics to cosmology. They have been used to model the
behavior of a variety of physical
systems including atmospheres\cite{NM1}, 
seismic activity of the earth\cite{NM2}, global ocean behavior\cite{NM3},
vibrations of crystals\cite{NM4}, molecules\cite{NM5}, and nuclei\cite{NM6}, 
functional motions of proteins, 
viruses and enzymes\cite{NM7}, the oscillation of rotating stars\cite{NM8},
gravitational wave response\cite{NM9}, black hole oscillations\cite{NM10},
liquids\cite{NM11}, ultracold trapped gases\cite{NM12}, and cold trapped ions
for quantum information processing\cite{NM13}.
Reflecting the ubiquitous appearance of small vibrations in nature, normal 
modes couple the complex motion of individual interacting particles into simple 
collective motions in which the particles move in sync with the same 
frequency and phase.
Systems in equilibrium experiencing small
perturbations tend to return to equilibrium if restorative forces are present. 
These restorative forces can often be 
approximated by effective harmonic terms that couple the $N$ particles in these
systems resulting in dynamics that can be transformed to that of $N$ 
 uncoupled oscillators
whose collective coordinates define the normal modes. The power of normal modes
lies in their ability to describe the complex motion of $N$ interacting 
particles in 
terms of collective coordinates whose character and 
frequencies reflect the 
inter-particle correlations of the system, thus incorporating many-body
effects into simple dynamic motions. Higher order effects can be 
expanded in this physically intuitive basis of normal modes. 
If higher order (e.g. anharmonic) effects are small,
these collective motions are eigenfunctions of an  
approximate Hamiltonian acquiring some measure of stability 
as a function of time; thus,
a system in a single normal
mode will have a tendency to remain in that mode until perturbed. Normal modes 
manifest the symmetry of this underlying approximate Hamiltonian
with the possibility of offering analytic solutions to a many-body problem 
and a clear
physical picture of the dynamics.

Confined quantum systems in the laboratory
 with $N$ identical interacting particles  
have been shown to exhibit collective behaviors thought to arise
 from general and powerful principles
of organization\cite{anderson,anderson2,guidry,laughlin,zaanen}. 
In a recent paper, collective behavior in the form of $N$-body
normal modes successfully described the thermodynamic behavior associated with
the superfluidity of an ultracold gas of fermions in the unitary 
regime\cite{emergence}. Two normal modes, selected by
the Pauli principle, were found to play a role in creating and stabilizing
the superfluid
 behavior at low temperatures, a phonon mode at ultralow temperatures
and a single particle radial excitation mode, i.e. a particle-hole excitation,
 as the temperature increases. 
This radial excitation has a much 
higher frequency  and creates a gap that stabilizes the superfluid behavior.
The two normal modes were found to describe the thermodynamic behavior of 
this gas quite
well compared to experimental data.

These normal modes are the perturbation solutions at first order in inverse
dimensionality of a first principle many-body formalism called
symmetry invariant perturbation theory (SPT).  This formalism  uses a
group theoretic approach for the solution of a fully interacting, 
 many-body, three-dimensional
Hamiltonian with an arbitrary interaction potential as well as a 
confining potential\cite{paperI,JMPpaper}.  Using the
symmetry of the symmetric group which can be accessed at large 
dimension\cite{FGpaper}, this approach has successfully rearranged
 the many-body work needed at each order in the perturbation
series so that an exact solution can, in principle, be obtained order-by-order 
using group theory and graphical techniques, i.e. 
non-numerically\cite{rearrangeprl}. 
Specifically, the numerical work has been rearranged
 into analytic building blocks that allow a formulation 
that does not scale with 
$N$\cite{JMPpaper, test, toth, rearrangeprl, complexity}.
Group theory is 
 used to partition the $N$ scaling problem away from the interaction 
dynamics, allowing the $N$ scaling to be treated as a  separate mathematical 
problem (cf. the Wigner-Eckart theorem). 
The exponential scaling is shifted from a dependence on the number of
particles, $N$, to a dependence on the order of the perturbation 
expansion\cite{complexity}. This allows one to obtain exact
first-order results that contain beyond-mean-field effects for 
all values of $N$ from a single calculation, but going to higher order
is now exponentially
difficult. 
 The analytic building blocks
have been calculated and stored to minimize the work needed
for new calculations\cite{epaps}. 
Since the perturbation does not involve the
strength of the interaction, strongly interacting systems can be studied. 

Initially applied to systems of cold 
bosons, my group has previously derived 
beyond-mean-field energies\cite{FGpaper,energy}, frequencies\cite{energy},
normal mode coordinates\cite{paperI}, wave functions\cite{paperI} and density 
profiles\cite{laingdensity}
for general isotropic, interacting confined quantum systems of identical 
bosons.
Recently I have extended 
this formalism
to systems of fermions\cite{prl,harmoniumpra,partition,emergence}.
I avoid the numerically demanding construction of explicitly 
antisymmetrized wave
functions by applying the Pauli principle at first order ``on paper'' through 
the assignment
of appropriate normal mode quanta\cite{prl,harmoniumpra}. I have determined
ground\cite{prl}
 and excited state\cite{emergence} beyond-mean-field
energies and their degeneracies 
  allowing the construction of a partition 
function\cite{partition}.

Analytic expressions for these normal modes have been obtained
in a previous paper, Ref.~\cite{paperI}.  In Sections 5 and 6 of 
Ref.~\cite{paperI}, we discussed the symmetry of the $N$-body
quantum-confinement problem at large dimension 
which greatly simplifies the problem, 
making possible, in principle, an exact solution of this
$N$-body problem, with $N(N-1)/2$ interparticle interactions order-by-order. 
In Section 7 we exploited
this symmetry, deriving symmetry coordinates used
in the determination of the normal
modes of the system. We introduced a
particular approach to derive a suitable basis of symmetry
coordinates that builds up the complexity slowly and systematically. This
is illustrated in detail for each of the five types of symmetry coordinates 
that transform
under the five different irreducible representations for a system of $N$ 
identical particles. In Section 8 we applied this
general theory to derive in detail analytic
expressions for
the normal-mode coordinates.

These functions serve as a
natural basis for the determination of higher order terms in the
perturbation series, and they offer the possibility of a clear physical
picture of the dynamics if higher order terms are small. One major advantage of
this approach is that $N$ appears as a parameter in the analytical
expressions for the normal modes, as well as the energy spectrum,  
so the behavior of these modes can be easily studied as a function of $N$.

The  quantum wave function yields important information about the dynamics of a
system beyond that obtainable from the energy spectrum. Although explicit
antisymmetrized wave functions are not currently obtained in this SPT
formalism, the normal mode solutions at first order constitute a complete basis.
Thus the character of the wave function may be revealed if a single normal
mode is dominant since the normal modes have clear, macroscopic motions. 

In this paper, I expand on our earlier discussion of these analytic normal
modes, examining in detail the motion of individual particles as they 
contribute to the
 five types of normal modes for an $N$-body system of identical particles under
quantum confinement. In 
particular, I study the evolution of collective behavior as a function of 
$N$ from few-body systems to many-body systems, 
first making general observations and then choosing as a specific
case the Hamiltonian for a confined system of fermions in the unitary regime
which is known to support collective behavior.
Finally in the Appendices, I present additional details of the derivation 
of these normal modes.

\section{Review of the N-body Normal Mode Derivation}

In this Section, I briefly review the derivation of the normal modes
that was presented in more detail in Ref.~\cite{paperI}
as solutions to the SPT first order perturbation equation.

\subsection{${\mathbf{D}}$-dimensional ${\mathbf{N}}$-body Schr\"odinger
Equation}\label{sec:SE} 

The  $D$-dimensional Schr\"odinger equation
in Cartesian coordinates for a system of $N$ particles 
interacting via
a two-body interaction potential $g_{ij}$, and 
confined by a spherically symmetric potential $V_{conf}$ is

\begin{equation} \label{generalH} 
H \Psi  =  \left[ \sum\limits_{i=1}^{N} h_{i} +
\sum_{i=1}^{N-1}\sum\limits_{j=i+1}^{N} g_{ij} \right] \Psi = E
\Psi \,,  
\end{equation} 

\begin{equation} \label{generalH1} 
\begin{array}{rcl}
h_{i} & = & -\frac{\hbar^2}{2
m_{i}}\sum\limits_{\nu=1}^{D}\frac{\partial^2}{\partial
x_{i\nu}^2} +
V_{\mathtt{conf}}\left(\sqrt{\sum\nolimits_{\nu=1}^{D}x_{i\nu}^2}\right)
\,,  \\
g_{ij} & = & V_{\mathtt{int}}\left(\sqrt{\sum\nolimits_{\nu=1}^{D}\left(x_{i\nu}-x_{j\nu}
\right)^2}\right),  
\end{array}
\end{equation}

%
\noindent where $h_{i}$ is the single-particle Hamiltonian and
 $x_{i\nu}$ is the $\nu^{th}$ Cartesian component
of the $i^{th}$ particle. 
Transforming the  Sch\"odinger equation from Cartesian to the internal 
coordinates is accomplished using: 
\begin{equation}\label{eq:int_coords}
\renewcommand{\arraystretch}{1.5}
\begin{array}{rcl} r_i & = &\sqrt{\sum_{\nu=1}^{D} x_{i\nu}^2}\,, \;\;\; (1 \le i \le
N)\,,
\;\;\; 
\\ \gamma_{ij} & = & cos(\theta_{ij})=\left(\sum_{\nu=1}^{D}
x_{i\nu}x_{j\nu}\right) / r_i r_j\,,
\end{array}
\renewcommand{\arraystretch}{1}
\end{equation}
\noindent $(1 \le i < j \le N)$\,, which are the $D$-dimensional scalar radii $r_i$ of the $N$
particles from the center of the confining potential and the
cosines $\gamma_{ij}$ of the $N(N-1)/2$ angles between the radial
vectors.

A similarity transformation removes the first-order
 derivatives, while the second derivative terms drop out
in the $D\to\infty$ limit, yielding a static zeroth-order
problem. First order corrections result in
 simple harmonic normal-mode oscillations about the
infinite-dimensional structure.

The Schr\"odinger equation 
becomes\cite{avery}, $ (T+V)\, \Phi = E \,\Phi$
where:
\begin{widetext}
\begin{equation}
T 
= 
{\displaystyle \hbar^2
\sum\limits_{i=1}^{N}\Biggl[-\frac{1}{2
m_i}\frac{\partial^2}{{\partial r_i}^2}- \frac{1}{2 m_i r_i^2}
\sum\limits_{j\not=i}\sum\limits_{k\not=i}
\frac{\partial}{\partial\gamma_{ij}}(\gamma_{jk}-\gamma_{ij}
\gamma_{ik})
\frac{\partial}{\partial\gamma_{ik}}}
{\displaystyle +\frac{N(N-2)+(D-N-1)^2 \left(
\frac{\Gamma^{(i)}}{\Gamma} \right) }{8 m_i r_i^2} \Biggr] }
 \label{eq:SE_T}
\end{equation}
\end{widetext}
\begin{equation}
V=\sum\limits_{i=1}^{N}V_{\mathtt{conf}}(r_i)+
\sum\limits_{i=1}^{N-1}\sum\limits_{j=i+1}^{N}
V_{\mathtt{int}}(r_{ij}) \,,
\end{equation}
%
$\Gamma$ is the Gramian determinant of the
matrix with elements $\gamma_{ij}$ (see
Appendix D in Ref~\cite{FGpaper}), and
$\Gamma^{(i)}$ is the determinant of the
$i^{th}$ principal minor where the row
and column of the $i^{th}$ particle have been deleted. The quantity
$r_{ij}=\sqrt{r_{i}^2+r_{j}^2-2r_{i}r_{j}\gamma_{ij}}$ is the
interparticle separation. 
 From
Eq.~(\ref{eq:SE_T}), it is clear that all first-order derivatives have been
eliminated from the Hamiltonian.

\subsection{Infinite-${\mathbf{D}}$ analysis: Zeroth-order
energy}\label{sec:infD}

Dimensionally scaled variables are defined: $\bar{r}_i = r_i/\kappa(D), \,\,
 \bar{E} = \kappa(D) E$ and $\bar{H} = \kappa(D)H$, where $\kappa(D)$ is
a  scale factor that regularizes
the large-dimension limit of the Schr\"odinger equation. The exact form
of $\kappa(D)$ is not fixed, but can be chosen to yield scaling results that are
as simple as possible while satisfying $\kappa(D) \sim D^2$.  
Examples of $\kappa(D)$
for different systems are given  after Eq.~(10) of Ref.~\cite{laingdensity}.

The
factor of $\kappa(D)$ plays the role of an effective mass that increases with $D$, suppressing the derivative terms but leaving  a
centrifugal-like term in an effective potential,
\begin{eqnarray}
\label{veff}
\bar{V}_{\mathtt{eff}}(\bar{r},\gamma;\delta=0)&=&\sum\limits_{i=1}^{N}\left(\frac{\hbar^2}{8
m_i
\bar{r}_i^2}\frac{\Gamma^{(i)}}{\Gamma}+\bar{V}_{\mathtt{conf}}(\bar{r},\gamma;\delta=0)\right)\nonumber \\
&& +\sum\limits_{i=1}^{N-1}\sum\limits_{j=i+1}^{N}
\bar{V}_{\mathtt{int}}(\bar{r},\gamma;\delta=0)\,,
\end{eqnarray}
where $\delta=1/D$, and the particles become frozen at large
$D$. 
We assume  all radii and angle cosines of the particles are equal when
$D\to\infty$, i.e.
$\bar{r}_{i}=\bar{r}_{\infty} \;\; (1 \le i \le N)$ and
$\gamma_{ij}=\overline{\gamma}_\infty \;\; (1 \le i < j \le N)$
where $\bar{r}_{\infty}$ and
$\overline{\gamma}_\infty$ satisfy:

\begin{equation}
\label{minimum1} \left[ \frac{\partial
\bar{V}_{\mathtt{eff}}(\bar{r},\gamma;\delta)}{\partial
\bar{r}_{i}} \right]_{\delta=0}=0,\,\,\,\,\,\,\,
 \left[ \frac{\partial \bar{V}_{\mathtt{eff}}(\bar{r},\gamma;\delta)}{\partial
\gamma_{ij}}\right]_{\delta=0}=0,
\end{equation}
%
resulting in a maximally symmetric structure.
In scaled units the zeroth-order
($D\to\infty$) approximation for the energy 
is: $\bar{E}_{\infty}=\bar{V}_{\mathtt{eff}}(\bar{r}_{\infty})$, while
the centrifugal-like term
in $\bar{V}_{\mathtt{eff}}$, which is nonzero even for the ground
state, is a zero-point energy contribution from the minimum
uncertainty principle\cite{chat}.

\subsection{The ${\mathbf{1/D}}$ first-order
energy correction}\label{sec:firstorder}

At zeroth-order, the particles can be viewed as frozen in a
maximally symmetric, high-$D$ configuration.
 Solving Eqs. (\ref{minimum1}) for $\bar{r}_{\infty}$ and
$\overline{\gamma}_\infty$ yields the infinite-$D$ structure and
zeroth-order energy providing the starting point for the $1/D$
expansion. In order to determine the $1/D$ quantum correction to the energy
for large but finite values of $D$, we expand about the minimum of
the $D\to\infty$ effective potential. A position
vector of the $N(N+1)/2$ internal coordinates is defined as:

\begin{equation}\label{eq:ytranspose}
\begin{array}[t]{l} {\bar{\bm{y}}} = \left( \begin{array}{c} \bar{\bm{r}} \\
\bm{\gamma} \end{array} \right) \,, \;\;\; \mbox{where} \;\;\; \\
\mbox{and} \;\;\; \bar{\bm{r}} = \left(
\begin{array}{c}
\bar{r}_1 \\
\bar{r}_2 \\
\vdots \\
\bar{r}_N
\end{array}
\right) \,. \end{array} 
\bm{\gamma} = \left(
\begin{array}{c}
\gamma_{12} \\ \cline{1-1}
\gamma_{13} \\
\gamma_{23} \\ \cline{1-1}
\gamma_{14} \\
\gamma_{24} \\
\gamma_{34} \\ \cline{1-1}
\gamma_{15} \\
\gamma_{25} \\
\vdots \\
\gamma_{N-2,N} \\
\gamma_{N-1,N} \end{array} \right) \,,
\end{equation}

\noindent The following substitutions are made for all
radii and angle cosines: $\bar{r}_{i} = \bar{r}_{\infty}+\delta^{1/2}\bar{r}'_{i}$
and $\gamma_{ij} =
\overline{\gamma}_{\infty}+\delta^{1/2}\overline{\gamma}'_{ij}$ and a
power series is obtained in $\delta^{1/2}$ for the effective potential about
the $D\to\infty$ symmetric minimum:
$\left[  \frac{\partial \bar{V}_{\mathtt{eff}}}{\partial
\bar{y}_{\mu}}  \right]_{\delta^{1/2}=0} = 0$.
 Defining a displacement vector consisting of the internal
displacement coordinates:

\begin{equation}\label{eq:ytransposeP}
\begin{array}[t]{l} {\bar{\bm{y}}'} = \left( \begin{array}{c}\bar{r}' \\
\overline{\bm{\gamma}}' \end{array} \right) \,, \;\;\;
\mbox{where} \;\;\; \\ \mbox{and} \;\;\; \bar{\bm{r}}' = \left(
\begin{array}{c}
\bar{r}'_1 \\
\bar{r}'_2 \\
\vdots \\
\bar{r}'_N
\end{array}
\right) \,, \end{array} 
\overline{\bm{\gamma}}' = \left(
\begin{array}{c}
\overline{\gamma}'_{12} \\ \cline{1-1}
\overline{\gamma}'_{13} \\
\overline{\gamma}'_{23} \\ \cline{1-1}
\overline{\gamma}'_{14} \\
\overline{\gamma}'_{24} \\
\overline{\gamma}'_{34} \\ \cline{1-1}
\overline{\gamma}'_{15} \\
\overline{\gamma}'_{25} \\
\vdots \\
\overline{\gamma}'_{N-2,N} \\
\overline{\gamma}'_{N-1,N} \end{array} \right) \,,
\end{equation}

\noindent the expression for $\bar{V}_{\mathtt{eff}}$ becomes:

\begin{eqnarray}
\lefteqn{\bar{V}_{\mathtt{eff}}({\bar{\bm{y}}'}; \hspace{1ex}
\delta) = \left[ \bar{V}_{\mathtt{eff}} \right]_{\delta^{1/2}=0} }
\nonumber \\ && + \frac{1}{2} \, \delta \left\{
\sum\limits_{\mu=1}^{P} \sum\limits_{\nu=1}^{P} \bar{y}'_{\mu}
\left[\frac{\partial^2 \bar{V}_{\mathtt{eff}}}{\partial
\bar{y}_{\mu}
\partial \bar{y}_{\nu}}\right]_{\delta^{1/2}=0} \hspace{-1.5em} \bar{y}'_{\nu} + v_o \right\} +
O\left(\delta^{3/2}\right) \,, \nonumber \\ \label{Taylor}
\end{eqnarray}

\noindent where $P =N(N+1)/2$ is the number of internal coordinates
and $v_o = \left[  \frac{\partial
\bar{V}_{\mathtt{eff}}}{\partial \delta}\right]_{\delta^{1/2}=0}$.
The derivative terms in the kinetic energy
have a similar series expansion:
\begin{equation}\label{eq:T}
{\mathcal T}=-\frac{1}{2} \delta \sum\limits_{\mu=1}^{P}
\sum\limits_{\nu=1}^{P} {G}_{\mu\nu}
\partial_{\bar{y}'_{\mu}}
\partial_{\bar{y}'_{\nu}} + O\left(\delta^{3/2}\right),
\end{equation}
where ${\mathcal T}$ is the derivative portion of the kinetic
energy $T$ (see Eq.~(\ref{eq:SE_T})).  Thus, determining the
energy at first order is reduced to a harmonic problem,
which is solved by obtaining the normal modes of the system.
From Eqs. (\ref{Taylor}) and (\ref{eq:T}),
 $\bm{G}$ and ${\bf F}$, both constant matrices, are defined
in the first-order $\delta=1/D$ Hamiltonian below:

\begin{equation} \label{eq:Gham}
\widehat{H}_1=-\frac{1}{2} {\partial_{\bar{y}'}}^{T} {\bm G}
{\partial_{\bar{y}'}} + \frac{1}{2} \bar{\bm{y}}^{\prime T} {\bm
F} {{\bar{\bm{y}}'}} + v_o \,.
\end{equation}

\subsection{FG Matrix Method for the Normal Mode Frequencies and Coordinates}
The FG matrix method\cite{dcw} is used to obtain the normal-mode
vibrations and the harmonic-order energy correction. A
review of the FG matrix method is presented in Appendix A of 
Ref.~\cite{paperI},
but a brief summary is given below.

The $b^{\rm th}$ normal
mode coordinate may be written as (Eq.~(A9) Ref.~\cite{paperI})
\begin{equation} \label{eq:qyt}
[{\bm q'}]_b = {\bm{b}}^T {\bar{\bm{y}}'} \,,
\end{equation}
where the coefficient vector ${\bm{b}}$ satisfies the
eigenvalue equation (Eq.~(A10) Ref.~\cite{paperI})
\begin{equation} \label{eq:FGit}
{\bf F} \, \bm{G} \, {\bm{b}} = \lambda_b \, {\bm{b}}
\end{equation}
with the resultant secular equation (Eq.~(A11) Ref.~\cite{paperI})
$\det({\bf F}\bm{G}-\lambda{\bf I})=0$\,. The coefficient vector
also satisfies the normalization condition (Eq.~(A12) Ref.~\cite{paperI})
${\bm{b}}^T \bm{G} \, {\bm{b}} = 1$\,, and the frequencies are:
$\lambda_b=\bar{\omega}_b^2$\, (Eq.~(A3) Ref.~\cite{paperI}).

In an earlier paper\cite{FGpaper}, we solve these equations for the
frequencies.
The number of roots
$\lambda$ is equal to $P \equiv
N(N+1)/2$. However, due to the $S_N$
symmetry (see Ref.~\cite{hamermesh} and Appendix~\ref{app:Char}),
 there is a 
simplification to only five
distinct roots, $\lambda_{\mu}$, where $\mu$ runs over ${\bf
0}^-$, ${\bf 0}^+$, ${\bf 1}^-$, ${\bf 1}^+$, and ${\bf 2}$,
(see Refs.~\cite{FGpaper, loeser}).
Thus the energy through first-order in $\delta=1/D$
 (see Eq.~(\ref{eq:E1})) can be
written in terms of the five distinct normal-mode vibrational
frequencies\cite{FGpaper}.

\begin{equation}
\overline{E} = \overline{E}_{\infty} + \delta \Biggl[
\sum_{\renewcommand{\arraystretch}{0}
\begin{array}[t]{r@{}l@{}c@{}l@{}l} \scriptstyle \mu = \{
  & \scriptstyle \bm{0}^\pm,\hspace{0.5ex}
  & \scriptstyle \bm{1}^\pm & , 
  &  \,\scriptstyle \bm{2}   \scriptstyle  \}
            \end{array}
            \renewcommand{\arraystretch}{1} }
(n_{\mu}+\frac{1}{2} d_{\mu})
\bar{\omega}_{\mu} \, + \, v_o \Biggr] \,. \label{eq:E1}
\end{equation}

\noindent where $\overline{E}_{\infty}$ is the energy minimum as 
$\delta \rightarrow0$,
 $n_{\mu}$ is the total number of quanta in the normal mode
with the frequency $\bar{\omega}_{\mu}$;
 $\mu$ is a label which
runs over the five types of normal modes, ${\bf 0}^-$\,, ${\bf
0}^+$\,, ${\bf 1}^-$\,, ${\bf 1}^+$\,, and ${\bf 2}$\,, (irrespective of
 the particle number, see Ref.~\cite{FGpaper}and Ref.[15]
in \cite{paperI}), and $v_o$ is a constant (defined above and in 
Ref.~\cite{FGpaper}, Eq. 125).
The
multiplicities of the five roots are:
$d_{{\bf 0}^+} = 1, \hspace{1ex} d_{{\bf 0}^-} = 1,\;
d_{{\bf 1}^+} = N-1,\;  d_{{\bf 1}^-} = N-1,\;
d_{{\bf 2}} = N(N-3)/2$.

\subsection{Symmetry of the $\bm{F}$, $\bm{G}$ and $\bm{FG}$ Matrices}
\label{sec:symm}

The large degeneracy of the
frequencies 
indicates a very high degree of symmetry which is manifested in
the $\bm{F}$\,, $\bm{G}$\,,
and $\bm{FG}$ matrices which 
are $P \times P$ matrices. The $S_N$ symmetry of these
matrices, whose elements are
evaluated  for the maximally symmetric structure at large dimension, allows
them to be written in terms of six simple submatrices which are
invariant under 
$S_N$\,(See Ref~\cite{FGpaper}). 
The number of $r_i$ coordinates is $N$ and the
number of $\gamma_{ij}$ coordinates is $N(N-1)/2$\,. 
 These
matrices are invariant under interchange of the particles, effected
by the point group $S_N$\cite{FGpaper}.

We can thus write the $\bm{F}$, $\bm{G}$ and $\bm{FG}$
matrices with the following structure:
\begin{eqnarray} 
{\bf F}&=&\left(\begin{array}{cc} {\bf F}_{\bar{\bm{r}}'
\bar{\bm{r}}'} & {\bf F}_{\bar{\bm{r}}' \overline{\bm{\gamma}}'}
\\ {\bf F}_{\overline{\bm{\gamma}}' \bar{\bm{r}}'} & {\bf
F}_{\overline{\bm{\gamma}}' \overline{\bm{\gamma}}'}
\end{array} \right) \,\,\,\,\,\, 
%
{\bf G}=\left(\begin{array}{cc} {\bf G}_{\bar{\bm{r}}'
\bar{\bm{r}}'} & {\bf G}_{\bar{\bm{r}}' \overline{\bm{\gamma}}'}
\\ {\bf G}_{\overline{\bm{\gamma}}' \bar{\bm{r}}'} & {\bf
G}_{\overline{\bm{\gamma}}' \overline{\bm{\gamma}}'} \\
\end{array} \right)  \label{eq:G} \\
%
{\bf FG}&=&\left(\begin{array}{cc} {\bf FG}_{\bar{\bm{r}}'
\bar{\bm{r}}'} & {\bf FG}_{\bar{\bm{r}}' \overline{\bm{\gamma}}'}
\\ {\bf FG}_{\overline{\bm{\gamma}}' \bar{\bm{r}}'} & {\bf
FG}_{\overline{\bm{\gamma}}' \overline{\bm{\gamma}}'}
\end{array} \right) \label{eq:FG}
\end{eqnarray}

The structure of these matrices results in highly degenerate eigenvalues
and causes a reduction from
a possible $P=N(N+1)/2$ distinct frequencies to just five distinct
frequencies for $L = 0$ systems.

\subsection{Symmetry Coordinates}
\label{subsec:symnorm}

The $\bm{FG}$ matrix is
invariant under $S_N$\,, so it does not connect subspaces
belonging to different irreducible representations of
$S_N$\cite{WDC}. Thus from Eqs.~(\ref{eq:qyt}) and (\ref{eq:FGit})
the normal coordinates must transform under irreducible
representations of $S_N$\,. The normal coordinates will be
linear combinations of the elements of the internal coordinate
displacement vectors $\bar{\bm{r}}'$ and
$\overline{\bm{\gamma}}'$ which  transform under reducible matrix 
representations of $S_N$\,,
each spanning the corresponding carrier spaces.
(Appendix~\ref{app:Char}).

The radial displacement coordinate $\bar{\bm{r}}'$ transforms under a
reducible representation that reduces to one $1$-dimensional
irreducible representation labelled by the partition $[N]$ (the
partition denotes a corresponding Young diagram 
 of an irreducible representation (see
Appendix~\ref{app:Char})) and one $(N-1)$-dimensional irreducible
representation labelled by the partition $[N-1, \hspace{1ex} 1]$.
The angular displacement coordinate $\overline{\bm{\gamma}}'$ transforms
under a reducible representation that reduces to one
$1$-dimensional irreducible representation labelled by the
partition $[N]$, one $(N-1)$-dimensional irreducible
representation labelled by the partition $[N-1, \hspace{1ex} 1]$
and one $N(N-3)/2$-dimensional irreducible representation labelled
by the partition $[N-2, \hspace{1ex} 2]$.

We define the symmetry coordinate vector, $S$ as:

\begin{equation}\label{eq:trial}
\bm{S} = \left( \begin{array}{l} {\bm{S}}_{\bar{\bm{r}}'}^{[N]} \\
{\bm{S}}_{\overline{\bm{\gamma}}'}^{[N]} \\
{\bm{S}}_{\bar{\bm{r}}'}^{[N-1, \hspace{1ex} 1]} \\
{\bm{S}}_{\overline{\bm{\gamma}}'}^{[N-1, \hspace{1ex} 1]} \\
{\bm{S}}_{\overline{\bm{\gamma}}'}^{[N-2, \hspace{1ex} 2]}
\end{array} \right) =
\left( \begin{array}{l}W_{\bar{\bm{r}}'}^{[N]} \, \bar{\bm{r}}' \\ 
W_{\overline{\bm{\gamma}}'}^{[N]} \, \overline{\bm{\gamma}}' \\
W_{\bar{\bm{r}}'}^{[N-1, \hspace{1ex} 1]} \bar{\bm{r}}'\\
W_{\overline{\bm{\gamma}}'}^{[N-1, \hspace{1ex} 1]} \,
\overline{\bm{\gamma}}' \\
W_{\overline{\bm{\gamma}}'}^{[N-2, \hspace{1ex} 2]} \,
\overline{\bm{\gamma}}' \end{array} \right) \,,
\end{equation}
where the $W_{\bar{\bm{r}}'}^{[\alpha]}$ and the $ W_{\bar{\bm{\gamma}}'}^{[\alpha]}$
are the transformation matrices. 
This is shown in Ref.~\cite{paperI} using the theory of group characters 
to decompose
$\bar{\bm{r}}'$ and $\overline{\bm{\gamma}}'$ into basis functions that 
transform under these five irreducible representations of $S_N$.

The process used in Ref.~\cite{paperI} to determine the symmetry coordinates, 
and hence the $W_{\bar{r}'}^{{\alpha}}$ and
$W_{\overline{\gamma}'}^{{\alpha}}$ matrices was chosen to 
ensure that the $W$ matrices
satisfy the orthogonality restrictions between different irreducible
representations.  
This process also ensured that the sets of
coordinates transforming irreducibly under $S_N$ have the
simplest functional forms possible. One of the symmetry
coordinates was chosen to describe the simplest motion possible under the
requirement that it transforms irreducibly under $S_N$. The
succeeding symmetry coordinate was then chosen to have the next simplest
possible functional form that transforms irreducibly
 under $S_N$ etc.  In this way the
complexity of the motions described by the symmetry coordinates was
minimized, building up slowly as more symmetry
coordinates of a given species were added as $N$ increased, with no
disruption of lower $N$ symmetry coordinates. This method of determining the
symmetry coordinate basis is not unique, but was chosen to minimize the 
complexity. 

\subsection{Symmetry Coordinates and Transformation Matrices}

The five transformation matrices and the symmetry coordinates for five
irreducible representations are:
\begin{equation} \label{eq:WNeqsqrtN1}
[W^{[N]}_{\bar{\bm{r}}'}]_i = \frac{1}{\sqrt{N}} \,
[{\bm{1}}_{\bar{\bm{r}}'}]_i \;,\;\;  \;\;\;
{\bm{S}}_{\bar{\bm{r}}'}^{[N]} = \frac{1}{\sqrt{N}} \,
\sum_{i'=1}^N \overline{r}'_{i'} \,.
\end{equation}
\begin{eqnarray} 
[W^{[N]}_{\overline{\bm{\gamma}}'}]_{ij}&=& \sqrt{\frac{2}{N(N-1)}}
\,\, [{\bm{1}}_{\overline{\bm{\gamma}}'}]_{ij} \label{eq:WNgeqsqrt2ontnm11} \\
\mbox{and}\,\,\,\,\,\,
{\bm{S}}_{\overline{\bm{\gamma}}'}^{[N]}&=&
\sqrt{\frac{2}{N(N-1)}} \,\,\, \sum_{j'=2}^N \sum_{i' < j'}
\overline{\gamma}'_{i'j'} \,. \label{eq:SNm0}
\end{eqnarray}
where $[{\bm{1}}_{\bar{\bm{r}}'}]_i= 1 \;\; \forall \;\; 1 \leq i \leq N
\;\;\; \mbox{and} \;\;\; [{\bm{1}}_{\overline{\bm{\gamma}}'}]_{ij}
= 1  \;\; \forall \;\; 1 \leq i,j \leq N \,$.
\begin{eqnarray} \label{eq:WNm1r}
[W^{[N-1, \hspace{1ex} 1]}_{\bar{\bm{r}}'}]_{\xi i}&=&
\frac{1}{\sqrt{\xi(\xi+1)}} \left( \sum_{m=1}^\xi \delta_{mi} - \xi
\delta_{\xi+1,\, i} \right) 
\end{eqnarray}
\begin{equation} \label{eq:SNm1}
[{\bm{S}}_{\bar{\bm{r}}'}^{[N-1, \hspace{1ex} 1]}]_\xi =
\frac{1}{\sqrt{\xi(\xi+1)}} \left( \sum_{k'=1}^\xi \bar{r}'_{k'} - \xi
\bar{r}'_{\xi+1} \right)\,,
\end{equation}
\noindent where $1 \leq \xi \leq N-1$ and $1 \leq i \leq N$\,.
\begin{widetext}
\begin{equation}  \label{eq:SNm2}
\begin{array}{rcl}
{[W^{[N-1, \hspace{1ex} 1]}_{\overline{\bm{\gamma}}'}]_{\xi,\,ij}}
&=& {\frac{1}{\sqrt{\xi(\xi+1)(N-2)}}} \, \bigg( \big( \Theta_{\xi-i+1}
\, [{\bm{1}}_{\bar{\bm{r}}'}]_j + \Theta_{\xi-j+1} \,
[{\bm{1}}_{\bar{\bm{r}}'}]_i \big)
- \xi \big( \delta_{\xi+1,\, i} \,
[{\bm{1}}_{\bar{\bm{r}}'}]_j + \delta_{\xi+1,\, j} \,
[{\bm{1}}_{\bar{\bm{r}}'}]_i \big) \bigg) \,,\\
{[{\bm{S}}_{\overline{\bm{\gamma}}'}^{[N-1, \hspace{1ex} 1]}]_\xi}
&=&  {\displaystyle \frac{1}{\sqrt{\xi(\xi+1)(N-2)}} \, \left( \left[
\sum_{l' = 2}^\xi \, \sum_{k'=1}^{l'-1} \overline{\gamma}'_{k'l'} +
\sum_{k' = 1}^\xi \, \sum_{l'=k'+1}^{N} \hspace{-1ex}
\overline{\gamma}'_{k'l'} \right] - \xi \left[ \sum_{k'=1}^\xi
\overline{\gamma}'_{k',\,\xi+1} + \sum_{l'=\xi+2}^N \hspace{-0.5ex}
\overline{\gamma}'_{\xi+1,\,l'} \right] \right) \,. }
\end{array} \renewcommand{\arraystretch}{1}
\end{equation}
\end{widetext}
\noindent where $1 \leq \xi \leq N-1$ and $1 \leq i < j \leq N$\,.
\begin{widetext}
\begin{equation} \label{eq:SNm3}
\begin{split}
{[W^{[N-2, \hspace{1ex}2]}_{\overline{\bm{\gamma}}'}]_{ij,\,mn}}
& =  \frac{1}{\sqrt{i(i+1)(j-3)(j-2)}} \,
\Bigl( (\Theta_{i-m+1} - i
\delta_{i+1,\,m})
(\Theta_{j-n} -(j-3)\delta_{jn}) +  
(\Theta_{i-n+1} - i \delta_{i+1,\,n})\\
& \times (\Theta_{j-m}-(j-3)\delta_{jm}) \Bigr)  \,  \\
{[{\bm{S}}_{\overline{\bm{\gamma}}'}^{[N-2, \hspace{1ex} 2]}]_{ij}}
& =   \frac{1}{\sqrt{i(i+1)(j-3)(j-2)}} \, \Bigl(
\vphantom{\sum_{k=1}^{[j'-1, i]_{min}} \hspace{-2ex}
\overline{\gamma}'_{kj'}}  
 \sum_{j'=2}^{j-1} \sum_{k=1}^{[j'-1, i]_{min}}
\hspace{-2ex} \overline{\gamma}'_{kj'} + \sum_{k=1}^{i-1}
\sum_{j'=k+1}^i
\overline{\gamma}'_{kj'} - (j-3) \sum_{k=1}^i \overline{\gamma}'_{kj}  \\ 
&  - i \sum_{k=1}^{i} {\overline{\gamma}}'_{k,(i+1)}
 - i \sum_{j'=i+2}^{j-1} {\overline{\gamma}}'_{(i+1),j'} + i (j-3)
{\overline{\gamma}}'_{(i+1),j}  \Bigr) \,,
\end{split}
\end{equation}
\end{widetext}
\noindent where $1 \leq i \leq j-2$ and $4 \leq j \leq N$ and 
$1 \leq m < n \leq N$\,. We define the Heaviside step function as:

\begin{equation} \label{eq:Heaviside}
\begin{array}{r@{\hspace{1ex}}l} {\displaystyle \Theta_{i-j+1} =
\sum_{m=1}^i \delta_{mj} } & =  1 \mbox{ when } j-i<1 \\
& = 0 \mbox{ when } j-i \geq 1 \,. \end{array}
\end{equation}

\subsection{Transformation to  Normal Mode Coordinates}
 The invariance of Eq. (13) under $S_N$ means that the F, G and FG matrices 
used to solve for the first-order energies and normal modes transform under
irreducible representations of $S_N$.  When the FG matrix is transformed
from the $\bar{\bm{r}}'$ and $\overline{\bm{\gamma}}'$ basis to
symmetry coordinates, the full $N(N+1)/2 \times N(N+1)/2$ matrix is reduced to
block diagonal form yielding one $2 \times 2$ block for the $[N]$ sector, $N-1$
identical $2 \times 2$ blocks for the $[N-1,1]$ sector and $N(N-3)/2$ identical
$1 \times 1$ blocks for the $[N-2,2]$ sector.  

In the $[N]$ and $[N-1,1]$ sectors, the $2 \times 2$ blocks allow the
$\bar{\bm{r}}'$ and $\overline{\bm{\gamma}}'$ symmetry coordinates to mix
in the
normal coordinates.  The $1 \times 1$ structure in the $[N-2,2]$ sector reflects the fact that the $[N-2,2]$ normal modes are entirely angular i.e. there are
no  $\bar{\bm{r}}'$ symmetry coordinates in this sector.

We applied the $\bm{FG}$ method using these symmetry
coordinates to determine the eigenvalues, 
$\lambda_\alpha = {\bar{\omega}_\alpha}^2$, 
frequencies, $\bar{\omega}_\alpha$ and normal modes, $\bm{q}^\prime$ of the
system:
\begin{equation} \label{eq:lambda12pm}
\lambda^\pm_\alpha = \frac{a_\alpha \pm \sqrt{b_\alpha^2 +
4\,c_\alpha }}{2} 
\end{equation}
for the $\alpha=[N]$ and $[N-1, \hspace{1ex} 1]$ sectors, where
\begin{eqnarray} \label{eq:abcalpha}
a_\alpha & = &
\protect[\bm{\sigma_\alpha^{FG}}\protect]_{\bar{\bm{r}}',\,\bar{\bm{r}}'}
+
\protect[\bm{\sigma_\alpha^{FG}}\protect]_{\overline{\bm{\gamma}}',\,\overline{\bm{\gamma}}'}
\nonumber \\
b_\alpha & = &
\protect[\bm{\sigma_\alpha^{FG}}\protect]_{\bar{\bm{r}}',\,\bar{\bm{r}}'}
-
\protect[\bm{\sigma_\alpha^{FG}}\protect]_{\overline{\bm{\gamma}}',\,\overline{\bm{\gamma}}'}
\\ c_\alpha & = & \protect[\bm{\sigma_\alpha^{FG}}\protect]_{\bar{\bm{r}}',\,
\overline{\bm{\gamma}}'} \, \times
\protect[\bm{\sigma_\alpha^{FG}}\protect]_{\overline{\bm{\gamma}}',\,\bar{\bm{r}}'}
\nonumber \,,
\end{eqnarray}
while $\lambda_{[N-2, \hspace{1ex} 2]} = \bm{\sigma_{[N-2,
\hspace{1ex} 2]}^{FG}}$\,. The $\bm{\sigma_\alpha^{FG}}$ are
the elements 
of the $\bm{FG}$ matrix of
Eq.~(\ref{eq:FG}) expressed in the basis of symmetry coordinates. 
The $\bm{\sigma_\alpha^{FG}}$ for
the $\alpha=[N]$ and $[N-1, \hspace{1ex} 1]$ sectors are $2 \times
2$ matrices (See Appendix B), while $\bm{\sigma_{[N-2, \hspace{1ex} 2]}^{FG}}$ is a
one-component quantity.  These quantities are defined generally 
in Ref.~\cite{paperI}
Eqs. (28, 29, 126, 162, 163) and also specifically 
in Ref.~\cite{FGpaper} for three different confining and
interparticle potentials.
The normal coordinates are:
\begin{equation} \label{eq:qnpfullexp}
{\bm{q}'}_{\pm}^{[N]}  = c_{\pm}^{[N]} \left(
\cos{\theta^{[N]}_{\pm}} \, [{\bm{S}}_{\bar{\bm{r}}'}^{[N]}]
\, + \, \sin{\theta^{[N]}_{\pm}} \,
[{\bm{S}}_{\overline{\bm{\gamma}}'}^{[N]}] \right)  
\end{equation}

\begin{equation} \label{eq:qnpfullexp1}
\begin{split}
{\bm{q}'}_{\xi\pm}^{[N-1,1]} & = c_{\pm}^{[N-1,1]} \Bigl(
\cos{\theta^{[N-1,1]}_{\pm}} [{\bm{S}}_{\bar{\bm{r}}'}^{[N-1,1]}]_{\xi}  \\
& \,\,\,\,\,\,\,\,\,\,\,\,\,\,\,\,\,\,\,\,\,
+ \sin{\theta^{[N-1,1]}_{\pm}}[{\bm{S}}_{\overline{\bm{\gamma}}'}^{[N-1,1]}]_{\xi}
 \Bigr) \, 
\end{split}
\end{equation}

\noindent for the $\alpha=[N]$ and $[N-1, \hspace{1ex} 1]$ sectors,
 $1 \leq \xi \leq N-1$ and
%

%
\begin{equation} \label{eq:qnm2fullexp}
{\bm{q}'}^{[N-2, \hspace{1ex} 2]} = c^{[N-2, \hspace{1ex} 2]}
{\bm{S}}_{\overline{\bm{\gamma}}'}^{[N-2, \hspace{1ex} 2]} \,
\end{equation}
for the $[N-2, \hspace{1ex} 2]$ sector.
The
$\bar{\bm{r}}'$-$\overline{\bm{\gamma}}'$ mixing angle,
$\theta^\alpha_\pm$\,, is given by
\begin{equation} \label{eq:tanthetaalphapm}
\tan{\theta^\alpha_\pm} = \frac{(\lambda^\pm_\alpha -
\protect[\bm{\sigma_\alpha^{FG}}\protect]_{\bar{\bm{r}}',\,\bar{\bm{r}}'})}
{\protect[\bm{\sigma_\alpha^{FG}}\protect]_{\bar{\bm{r}}',\,
\overline{\bm{\gamma}}'}} =
\frac{\protect[\bm{\sigma_\alpha^{FG}}\protect]_{\overline{\bm{\gamma}}',\,\bar{\bm{r}}'}}{(\lambda^\pm_\alpha
-
\protect[\bm{\sigma_\alpha^{FG}}\protect]_{\overline{\bm{\gamma}}',\,\overline{\bm{\gamma}}'})}\,,
\end{equation}
while the normalization constants $c^{[\alpha]}$ are given by
\begin{equation} \label{eq:calphapm}
\begin{array}{l}
c_\pm^{[\alpha]}   = {\displaystyle  \frac{1}{\sqrt{\left(
\begin{array}{c} \cos{\theta^{[\alpha]}_\pm} \\
\sin{\theta^{[\alpha]}_\pm} \end{array} \right)^T
\bm{\sigma_{\alpha}^{G}} \left( \begin{array}{c}
\cos{\theta^{[\alpha]}_\pm} \\ \sin{\theta^{[\alpha]}_\pm} \end{array}
\right)}}} 
\end{array}
\end{equation}

\begin{equation} \label{eq:calpha2}
{\displaystyle c^{[N-2, \hspace{1ex} 2]} =
\frac{1}{\sqrt{\bm{\sigma_{[N-2, \hspace{1ex} 2]}^G}}} } \,\,.
\end{equation}
The $\bm{\sigma_{\alpha}^{G}}$ are related to the elements of
the $\bm{G}$ matrix of
Eq.~(\ref{eq:Gham}).
One determines the  $\bar{\bm{r}}'$-$\overline{\bm{\gamma}}'$ mixing angles,
$\theta^{[\alpha]}_\pm$ for the $[N]$ and $[N-1,1]$ species  from
Eq.~(\ref{eq:tanthetaalphapm}). The normalization constants
$c^{[\alpha]}$ of Eqs.~(\ref{eq:qnpfullexp}) and
(\ref{eq:qnm2fullexp}) are determined from Eqs.~(\ref{eq:tanthetaalphapm})
(\ref{eq:calphapm}) and (\ref{eq:calpha2}). The normal mode
vector, ${\bm{q}'}$\,, is then determined through Eqs.~(\ref{eq:qnpfullexp})
 and (\ref{eq:qnm2fullexp}). The analytic normal coordinates for $N$ identical
particles are:
\begin{widetext}
\begin{equation}
\begin{split}
q_\pm^{\prime \, [N] } & =  c^{[N]}_\pm \cos{\theta^{[N]}_\pm}
\frac{1}{\sqrt{N}} \, \sum_{i'=1}^N \overline{r}'_{i'} 
+ c^{[N]}_\pm
\sin{\theta^{[N]}_\pm} \sqrt{\frac{2}{N(N-1)}} \,\,\,
\sum_{j'=2}^N \sum_{i' < j'} \overline{\gamma}'_{i'j'}, \\
[{\bm{q}'}_\pm^{[N-1, \hspace{1ex} 1]}]_\xi 
& =  c^{[N-1, \hspace{1ex} 1]}_\pm 
\cos{\theta^{[N-1, \hspace{1ex}
1]}_\pm} \frac{1}{\sqrt{\xi(\xi+1)}} \left( \sum_{k'=1}^\xi 
\bar{r}'_{k'} - \xi\bar{r}'_{\xi+1} \right)  
 + c^{[N-1, \hspace{1ex} 1]}_\pm \sin{\theta^{[N-1, \hspace{1ex}
1]}_\pm} \frac{1}{\sqrt{\xi(\xi+1)(N-2)}}  \nonumber \\
& \left( \left[
\sum_{l' = 2}^\xi \, \sum_{k'=1}^{l'-1} \hspace{-1ex}
\overline{\gamma}'_{k'l'} + \sum_{k' = 1}^\xi \, \sum_{l'=k'+1}^{N}
\hspace{-1ex} \overline{\gamma}'_{k'l'} \right]  
 -\xi  \left[ \sum_{k' = 1}^\xi \, \overline{\gamma}'_{k', \xi+1} +
\sum_{l'=\xi+2}^N \,\hspace{-1ex} \overline{\gamma}'_{\xi+1,l'} \right] \right) \\ 
\mbox{where}\,\,\, 1 \leq \xi \leq N-1, \\
[{\bm{q}'}^{[N-2, \hspace{1ex} 2]}]_{ij} 
& =  c^{[N-2, \hspace{1ex} 2]} \frac{1}{\sqrt{i(i+1)(j-3)(j-2)}} \, \Bigl(
\vphantom{\sum_{k=1}^{[j'-1, i]_{min}} \hspace{-2ex}
\overline{\gamma}'_{kj'}}  
 \sum_{j'=2}^{j-1} \sum_{k=1}^{[j'-1, i]_{min}}
\hspace{-2ex} \overline{\gamma}'_{kj'} + \sum_{k=1}^{i-1}
\sum_{j'=k+1}^i
\overline{\gamma}'_{kj'} - (j-3) \sum_{k=1}^i \overline{\gamma}'_{kj}  \\ 
&  - i \sum_{k=1}^{i} {\overline{\gamma}}'_{k,(i+1)}
 - i \sum_{j'=i+2}^{j-1} {\overline{\gamma}}'_{(i+1),j'} + i (j-3)
{\overline{\gamma}}'_{(i+1),j}  \Bigr) \,,
\end{split}
\end{equation}
\end{widetext}

\noindent where $1 \leq i \leq j-2$ and $4 \leq j \leq N$

\section{Motions Associated with the Symmetry Coordinates}
\label{sec:symmetrymotions}

In this section, I analyze the motions of the five types of symmetry 
coordinates as expressed in Eqs.~(\ref{eq:WNeqsqrtN1},
\ref{eq:SNm0}, \ref{eq:SNm1} - \ref{eq:SNm3}).
 For symmetry coordinates,
there is no mixing
of radial and angular motion, so
the motion is either totally radial or totally angular.  
The symmetry coordinates transform irreducibly under $S_N$ and 
result in a block diagonal form for the $H_0$ matrix. When these blocks
are diagonalized we obtain the normal coordinates which are the solutions
to the first order equation. For the $[N]$
and $[N-1,1]$ sectors which are found in both the 
radial and angular
decompositions, there is mixing of these radial and angular
symmetry coordinates in the normal
modes.  For the $[N-2,2]$ sector, there is no radial part, only an
angular part, so no mixing; the
symmetry coordinates are the normal coordinates apart from a normalization
constant.

The symmetry coordinates describe motion that is collective  with 
the particles participating in synchronized motion, i.e. moving with the
same frequency and phase. 
Since the symmetry coordinates, except for the $[N-2,2]$ modes, are not solutions
to the Hamiltonian at first order, they do not necessarily exhibit the motion of
particles governed by the Hamiltonian.  Their motions could be mixed 
significantly with
another symmetry coordinate of the same species 
to form a normal coordinate, a solution to
the Hamiltonian at first order.  I will analyze the motion of the 
symmetry coordinates first and then use the knowledge of these motions to 
understand the normal coordinate behavior.

I am interested in the motion of individual particles as they participate
in the collective synchronized motion of these symmetry coordinates.
 To determine the
motion of an individual particle,  I need
to back transform from the known functional form of a particular
symmetry coordinate to the scaled internal displacement coordinates, 
${\bar{r}'_i}$ and ${\overline{\gamma}'_{ij}}$ and then transform from 
the scaled to the unscaled
displacement coordinates to be able to visualize these displacements.
Using Eq.(19) I can obtain the dimensionally scaled 
$\bar{\bm{r}}'$ and $\overline{\gamma}'$ vectors by back 
transforming with the transpose of the $W$ matrices.  These dimensionally 
scaled variables can be transformed to the unscaled internal displacement
coordinates using
$\bar{r}_{i} = \bar{r}_{\infty}+\delta^{1/2}\bar{r}'_{i}$,
 $\gamma_{ij} = \overline{\gamma}_{\infty}+\delta^{1/2}\overline{\gamma}'_{ij}$
and  $ r_i= \kappa(D) \bar{r}_i$. The unscaled internal 
coordinates, $r_i$ and $\gamma_{ij}$, allow one to determine the radial 
distance
from the confinement center
and the interparticle angle of each pair of particles using 
$\gamma_{ij}=\cos \theta_{ij}$ and 
$\overline{\gamma}_{\infty}=\cos \theta_{\infty}$, so 
$\theta_{ij}=\arccos{\gamma_{ij}}$ and
$\theta_{\infty} = \arccos{\overline{\gamma}_{\infty}}$.
Then $r_i - r_{\infty}$ and $\theta_{ij}-\theta_{\infty}$ give displacements
from the
maximally symmetric zeroth-order configuration ($r_\infty, \gamma_\infty$)
that are easy to visualize, connecting to our physical intuition and thus 
contributing to our understanding
of how the motion of $N$ particles becomes collective.

\paragraph{Motions Associated with Symmetry
Coordinate ${\bm{S}}_{\bar{\bm{r}}'}^{[N]}$.} 
 The simplest
collective motion for a system of identical particles
occurs when every particle executes the same motion with
the same phase. This type of collective motion occurs for the symmetry 
coordinates of the $[N]$ modes, both the
radial symmetry coordinate ${\bm{S}}_{\bar{\bm{r}}'}^{[N]}$ and 
angular symmetry coordinate ${\bm{S}}_{\overline{\bm{\gamma}}'}^{[N]}$. There is
just one symmetry coordinate in each $[N]$ sector. 
I am interested in the unscaled displacement quantity 
${r'}_i =  \kappa(D) {\bar{r}'}_i 
= D^2 {\bar{a}}_{ho} {\bar{r}'}_i$.
For the $[N]$ symmetry coordinate
${\bm{S}}_{\bar{\bm{r}}'}^{[N]}$, ${\bar{r}'}_i $ is obtained by back transforming
with $W_{\bar{r}'}^{[N]}$. Using
Eqs.~(\ref{eq:trial}) and (\ref{eq:WNeqsqrtN1}) the
motions associated with symmetry coordinate
${\bm{S}}_{\bar{\bm{r}}'}^{[N]}$ in the unscaled internal
displacement coordinates ${\bm{r}'}$ about the unscaled zeroth-order
configuration $r_\infty$
are given by:
\begin{eqnarray}
{\bm{r}'}^{[N]}&=& \overline{a}_{ho} \, D^{2} \,
\bar{\bm{r}}^{\prime [N]} = \overline{a}_{ho} \, D^{2} \,
[(W_{\bar{\bm{r}}'}^{[N]})]^T\,   {\bm{S}}_{\bar{\bm{r}}'}^{[N]} \, \\
&=& \overline{a}_{ho} \, 
\frac{D^2}{\sqrt{N}}
{\bm{S}}_{\bar{\bm{r}}'}^{[N]} \, {\bm{1}}_{\bar{\bm{r}}'} \,.\label{eq:rN}
\end{eqnarray}
The vector ${\bm{r}'}^{[N]}$ is a $N \times 1$ vector of the 
unscaled radial displacement coordinates
for all the particles participating in this collective motion. The motions
of all the particles are thus identical in this
 symmetry coordinate
${\bm{S}}_{\bar{\bm{r}}'}^{[N]}$ involving identical radial motions out and
then back in from
the positions of the zeroth-order configuration.  This results 
in a symmetric stretch collective motion, where all the radii
expand and contract together with decreasing amplitudes as $N$ increases. 
For $N=3$, a good molecular comparison is the stretching $A_1$ mode of
ammonia.

\smallskip

\paragraph{Motions Associated with Symmetry
Coordinate ${\bm{S}}_{\overline{\bm{\gamma}}'}^{[N]}$\,.}

Using Eqs.~(\ref{eq:trial})
and (\ref{eq:WNgeqsqrt2ontnm11}), the motions associated with
symmetry coordinate ${\bm{S}}_{\overline{\bm{\gamma}}'}^{[N]}$ in
the unscaled internal displacement coordinates ${\bm{\gamma}'}$ about
the zeroth-order configuration ${\bm{\gamma}}_\infty$
are
given by
\begin{eqnarray} 
{\bm{\gamma}'}^{[N]}&=&
 [(W_{\overline{\bm{\gamma}}'}^{[N]})]^T \,
{\bm{S}}_{\overline{\bm{\gamma}}'}^{[N]} \,\\
&=& \sqrt{\frac{2}{N(N-1)\, }} \,
{\bm{S}}_{\overline{\bm{\gamma}}'}^{[N]} \,
{\bm{1}}_{\overline{\bm{\gamma}}'} \,. \label{eq:SgammaN}
\end{eqnarray}

This vector ${\bm{\gamma}'}^{[N]}$ is a $N(N-1)/2 \times 1$ vector
 of the displacement contributions to the angle cosines 
for all the particles participating in this collective motion. The motions
of all the particles are thus identical in this
 symmetry coordinate
${\bm{S}}_{\bar{\bm{\gamma}}'}^{[N]}$ involving identical angular motions for
each pair of particles from the interparticle angles of the 
zeroth-order configuration.  This results 
in a symmetric bend collective motion, where all of the interparticle angles
expand and contract together with the radii unchanged. For $N=3$, a good
molecular comparison is the bending $A_1$ mode of ammonia.  As $N$ increases
this symmetric bending motion evolves into a center of mass motion with
small angular displacements for every interparticle angle while the radii
remain fixed. (See Section~\ref{subsec:general} for more detail.)

\paragraph{Motions Associated with Symmetry
Coordinates $[{\bm{S}}_{\bar{\bm{r}}'}^{[N-1, \hspace{1ex}
1]}]_\xi$\,.} Using Eqs.~(\ref{eq:trial}),
(\ref{eq:WNm1r}), and (\ref{eq:Heaviside}) the motions associated with
symmetry coordinates 
$[{\bm{S}}_{\bar{\bm{r}}'}^{[N-1,\hspace{1ex} 1]}]_{\xi}$ 
in the unscaled internal displacement
coordinates ${\bm{r}'}$ about the unscaled zeroth-order
configuration $\bm{r}_\infty$
are given by
\begin{equation} \label{eq:rNm1inr}
\begin{array}{r@{\hspace{0.5em}}c@{\hspace{0.5em}}l}
(r^{\prime [N-1, \hspace{1ex} 1]}_\xi)_i & = & \overline{a}_{ho} \,
D^{2} \, ({\bar{r}}^{\prime [N-1, \hspace{1ex} 1]}_{\xi})_i \\
&=& \overline{a}_{ho} \, D^{2} \, [{\bm{S}}_{\bar{\bm{r}}'}^{[N-1,
\hspace{1ex} 1]}]_\xi \,
[(W_{\bar{\bm{r}}'}^{[N-1, \hspace{1ex} 1]})_{\xi}]_i  \\
[1.5em] &=& {\displaystyle \overline{a}_{ho} \,
\frac{D^2}{\sqrt{\xi(\xi+1)}} [{\bm{S}}_{\bar{\bm{r}}'}^{[N-1,
\hspace{1ex} 1]}]_{\xi}} \, \\
&&\times \left( \Theta_{\xi-i+1} - \xi
\delta_{\xi+1,\, i} \right)
 \,.  
\end{array}
\end{equation}
The above equation gives the radial motion of the $i^{th}$
 particle participating 
in the collective motion of the ${\xi}^{th}$ symmetry coordinate in the radial
$[N-1,1]$ sector.  In this sector there are $N$-1 radial symmetry coordinates
i.e. $1 \leq \xi \leq N-1$, and the ${\xi}^{th}$ symmetry coordinate involves
the motion of the first $\xi +1$ particles.
(If $i>\xi+1$, the Heaviside and Kronecker delta functions are zero in 
Eq.~\ref{eq:rNm1inr}).
Thus the motion associated with symmetry coordinate
$[{\bm{S}}_{\bar{\bm{r}}'}^{[N-1, \hspace{1ex} 1]}]_1$ is an
antisymmetric stretch about the zeroth-order
configuration involving particles $1$ and $2$\,. As $\xi$ gets
larger, the motion involves more particles, $\xi+1$ particles,
with the first $\xi$ particles moving one way while the
$(\xi+1)^{\rm th}$ particle moves the other way. As
$\xi$ increases, the character of the motion evolves from an antisymmetric
stretch motion (a good molecular equivalent is an $E$ mode of ammonia),
to behavior that becomes more 
single-particle-like, i.e. a particle-hole excitation associated with radial
motion, since
the $(\xi+1)^{\rm th}$ radius vector in Eq.~(\ref{eq:rNm1inr}) is
weighted by the quantity $\xi$\,. (I examine this dependence on the
number of particles as $N$ increases in more detail in 
Section ~\ref{subsec:general}.)

\paragraph{Motions Associated with Symmetry
Coordinates $[{\bm{S}}_{\overline{\bm{\gamma}}'}^{[N-1, \hspace{1ex}
1]}]_\xi$\,.} Using Eqs.~(\ref{eq:trial}) and
(\ref{eq:SNm2}), the motions associated with symmetry coordinates
$[{\bm{S}}_{\overline{\bm{\gamma}}'}^{[N-1, \hspace{1ex} 1]}]_\xi$
in the unscaled internal displacement coordinates $\bm{\gamma}'$
about the unscaled zeroth-order configuration
$\bm{\gamma}_\infty = \gamma_\infty{\bm{1}}_{\overline{\bm{\gamma}}'}$ are given by:

\begin{widetext}
\begin{equation} \label{eq:gNm1ing}
\begin{split}
(\gamma^{\prime [N-1, \hspace{1ex} 1]}_\xi)_{ij}
& = 
 [{\bm{S}}_{\overline{\bm{\gamma}}'}^{[N-1,\hspace{1ex}
1]}]_{\xi} \, [(W_{\overline{\bm{\gamma}}'}^{[N-1, \hspace{1ex} 1]})_\xi]_{ij}   
  =  \frac{1}{\sqrt{\xi(\xi+1)(N-2)}} \, 
[{\bm{S}}_{\overline{\bm{\gamma}}'}^{[N-1,\hspace{1ex} 1]}]_\xi \,\\
& \times \bigg( \big( \Theta_{\xi-i+1} \,
[{\bm{1}}_{\overline{\bm{\gamma}}'}]_{ij} + \Theta_{\xi-j+1} \,
[{\bm{1}}_{\overline{\bm{\gamma}}'}]_{ij} \big)
 - \xi \big(
\delta_{\xi+1,\, i} \, [{\bm{1}}_{\overline{\bm{\gamma}}'}]_{ij} +
\delta_{\xi+1,\, j} \, [{\bm{1}}_{\overline{\bm{\gamma}}'}]_{ij}
\big) \bigg)
 \,.  
\end{split}
\end{equation}
\end{widetext}
The above equation gives the angular displacement of the ${i}^{th}$ and
${j}^{th}$ particles participating in the collective motion of the $\xi$
symmetry coordinate in the angular $[N-1,1]$ sector.
The ${\xi}^{th}$ symmetry coordinate in this angular sector $[N-1,1]$ involves
the angular motion of the first $\xi+1$ particles, thus affecting any angular
displacement ${\gamma}_{ij}$ where $1 \leq i \leq \xi+1$ or
$1 \leq j \leq \xi+1$.
All other angular displacements are zero. 
(When $i,j > \xi+1$, the Heaviside and Kronecker delta functions are zero.
The $\gamma_{12}$ displacement is
zero by cancellation.) Thus the motion associated with symmetry coordinate
$[{\bm{S}}_{\overline{\bm{\gamma}}'}^{[N-1, \hspace{1ex} 1]}]_1$
is an antisymmetric bending about the zeroth-order
configuration where the angle cosines $\gamma'_{13}$,
$\gamma'_{14}$, $\gamma'_{15}$, $\ldots$ increase, $\gamma'_{23}$,
$\gamma'_{24}$, $\gamma'_{25}$, $\ldots$ decrease while
$\gamma'_{12}$, $\gamma'_{34}$, $\gamma'_{35}$, $\gamma'_{45}$, $\ldots$
remain unchanged. Thus, analogously to the $\bar{\bm{r}}'$ sector of the $[N-1,
\hspace{1ex} 1]$ species,
$[{\bm{S}}_{\overline{\bm{\gamma}}'}^{[N-1, \hspace{1ex} 1]}]_1$
involves the motions of particles $1$ and $2$ moving with opposite
phase to each other. As $\xi$ gets
larger, the angular motion involves more particles, $\xi+1$ particles,
with the $(\xi+1)^{\rm th}$ particle moving with opposite phase to
the first $\xi$ particles. Analogously to
the radial sector of the $[N-1, \hspace{1ex} 1]$ species,
as $\xi$ increases the motion evolves from an antisymmetric stretch 
motion (cf. an $E$ mode of ammonia), to behavior that becomes
 more single-particle-like, i.e. a particle-hole excitation due to angular
displacement,
since the angle cosines involving the $(\xi+1)^{\rm th}$ particle
in Eq.~(\ref{eq:gNm1ing}) are weighted by the quantity $\xi$\,.

\smallskip

\paragraph{Motions Associated with Symmetry
Coordinate $[{\bm{S}}_{\overline{\bm{\gamma}}'}^{[N-2, \hspace{1ex}
2]}]_{ij}$\,.} Using Eqs.~(\ref{eq:trial}) and
(\ref{eq:SNm3}), the motions of the unscaled internal
displacement coordinates $\bm{\gamma}'$  about the unscaled
zeroth-order configuration  
$\bm{\gamma}_\infty = \gamma_\infty{\bm{1}}_{\overline{\bm{\gamma}}'}$
associated with the symmetry coordinates
$[{\bm{S}}_{\overline{\bm{\gamma}}'}^{[N-2, \hspace{1ex}
2]}]_{ij}$ are given by:
\begin{widetext}
\begin{equation}\label{eq:nm2g}
\begin{split}
(\gamma^{\prime [N-2, \hspace{1ex} 2]}_{ij})_{mn} =
& [{\bm{S}}_{\overline{\bm{\gamma}}'}^{[N-2,
\hspace{1ex}
2]}]_{ij} \, [W_{\overline{\bm{\gamma}}'}^{[N-2, \hspace{1ex} 2]}]_{ij,\,mn}   
 =    \frac{1}{\sqrt{i(i+1)(j-3)(j-2)}} \,
\, \, [{\bm{S}}_{\overline{\bm{\gamma}}'}^{[N-2, \hspace{1ex}
2]}]_{ij} \\
& \times \Bigl(  (\Theta_{i-m+1} - i
\delta_{i+1,\,m})(\Theta_{j-n} -(j-3)\delta_{jn}) + 
(\Theta_{i-n+1} - i \delta_{i+1,\,n})(\Theta_{j-m}
-(j-3)\delta_{jm}) \Bigr)  \,,  
\end{split}
\end{equation}
\end{widetext}
where $1 \leq m < n \leq N$, and $1 \leq i \leq j-2$ and $4 \leq j
\leq N$\,. 
The above equation gives the angular displacement (contribution to the
angle cosine) of the ${m}^{th}$ and
${n}^{th}$ particles participating in the collective motion of the
${ij}^{th}$ symmetry coordinate in the $[N-2,2]$ sector.  There are
$N(N-3)/2$ symmetry coordinates 
$[{\bm{S}}_{\overline{\bm{\gamma}}'}^{[N-2,
\hspace{1ex} 2]}]_{ij}$ in this sector labeled by 
$[{\bm{S}}_{\overline{\bm{\gamma}}'}^{[N-2,
\hspace{1ex} 2]}]_{14}$, $[{\bm{S}}_{\overline{\bm{\gamma}}'}^{[N-2,
\hspace{1ex} 2]}]_{24}$, 
$[{\bm{S}}_{\overline{\bm{\gamma}}'}^{[N-2,
\hspace{1ex} 2]}]_{15}$, $[{\bm{S}}_{\overline{\bm{\gamma}}'}^{[N-2,
\hspace{1ex} 2]}]_{25}$, $[{\bm{S}}_{\overline{\bm{\gamma}}'}^{[N-2,
\hspace{1ex} 2]}]_{35}$,
$[{\bm{S}}_{\overline{\bm{\gamma}}'}^{[N-2,\hspace{1ex} 2]}]_{16}$, 
$[{\bm{S}}_{\overline{\bm{\gamma}}'}^{[N-2,\hspace{1ex} 2]}]_{26}$, 
$[{\bm{S}}_{\overline{\bm{\gamma}}'}^{[N-2,\hspace{1ex} 2]}]_{36}$, 
$[{\bm{S}}_{\overline{\bm{\gamma}}'}^{[N-2,\hspace{1ex} 2]}]_{46}$,
$[{\bm{S}}_{\overline{\bm{\gamma}}'}^{[N-2,\hspace{1ex} 2]}]_{17}$,
$\ldots$
$[{\bm{S}}_{\overline{\bm{\gamma}}'}^{[N-2,\hspace{1ex} 2]}]_{N-2,N}$.

From Eq.~(\ref{eq:nm2g}), the symmetry coordinate
$[{\bm{S}}_{\overline{\bm{\gamma}}'}^{[N-2,\hspace{1ex} 2]}]_{ij}$ 
only involves motions
of the first j particles (Note $i \leq j-2$, and for 
the particle labels
$m,n > j$ the
Heaviside and Kronecker delta functions are zero). 
Thus, the complexity of the functional form
of $[{\bm{S}}_{\overline{\bm{\gamma}}'}^{[N-2, \hspace{1ex}
2]}]_{ij}$ and the motions it describes builds up slowly and
systematically as more particles are added to the system.
The symmetry coordinate, $[{\bm{S}}_{\overline{\bm{\gamma}}'}^{[N-2,
\hspace{1ex} 2]}]_{14}$ , involves the motion of only the first four
particles with the simultaneous opening of two interparticle angles
$\theta_{13}$ and $\theta_{24}$, 
and closing of two different interparticle 
angles, $\theta_{14}$ and $\theta_{23}$. 
(cf. the $E$ mode of methane. Note that there are no 
$[N-2, \hspace{1ex} 2]$ modes when $N$
drops below four.)
For this lowest value of $N$, the positive and negative angular displacements
have equal values, however as $N$ increases, this collective motion quickly
evolves to create a compressional motion with a single dominant 
angle opening and closing
 while the other interparticle angles that number $N(N-1)/2-1$ make small
adjustments. (See Section ~\ref{subsec:general} for a more detailed discussion.)

\section{Motions Associated with the Normal Modes}
\label{normalmodemotions}
From Eq.~(\ref{eq:qnpfullexp}) shown again below, the symmetry 
coordinates 
in the 
$[N]$ and $[N-1,1]$ sectors are mixed to form a normal coordinate.

\begin{equation} \label{eq:mixing}
\begin{split}
{\bm{q}'}_{\pm}^{[N]} & = c_\pm^{[N]} \left(
\cos{\theta^{[N]}_\pm} \, [{\bm{S}}_{\bar{\bm{r}}'}^{[N]}]
\, + \, \sin{\theta^{[N]}_\pm} \,
[{\bm{S}}_{\overline{\bm{\gamma}}'}^{[N]}] \right)  \\
{\bm{q}'}_{\xi\pm}^{[N-1,1]} & = c_{\pm}^{[N-1,1]} \Bigl(
\cos{\theta^{[N-1,1]}_{\pm}}[{\bm{S}}_{\bar{\bm{r}}'}^{[N-1,1]}]_{\xi}  \\
& \,\,\,\,\,\,\,\,\,\,\,\,\,\,\,\,\,\,  + \sin{\theta^{[N-1,1]}_{\pm}}
[{\bm{S}}_{\overline{\bm{\gamma}}'}^{[N-1,1]}]_{\xi} \Bigr) \,
\end{split} 
\end{equation}

\noindent where $1 \leq \xi \leq N-1$.
Thus, depending on the value of the mixing angles, the normal modes,
which are the  solutions at first order of the
Schr\"odinger equation will be
a mixture of radial and angular behavior for the $[N]$ and 
$[N-1,1]$ sectors.
The $[N-2,2]$ sector has only angular behavior as noted above and so does
not mix with other symmetry coordinates. The value of the mixing angles,
of course, depends on the Hamiltonian terms at this first perturbation
order. Choosing various confining and interparticle potentials will result
in different values for the mixing coefficients and different collective
motion as dictated by the Hamiltonian.

\section{Collective behavior as a function of $N$}
\label{sec:collective}

In this section, I consider the evolution of behavior of the 
symmetry coordinates and then the normal coordinates
as a function of the number of particles, $N$.  For low values of $N$ it is
possible to find molecular equivalents to characterize the behavior of the
symmetry coordinates.  
(These comparisons are of course only approximate since the molecular
systems have Hamiltonians with the hetero atom providing Coulombic
confinement.)
For the [N] radial and angular modes,
the $A_1$ modes of ammonia, $NH_3$ with their symmetric breathing and bending
motions are good equivalents. The radial and angular [N-1,1] modes
are similar to the two $E$ ammonia modes for small $N$.  These motions show one
of the hydrogen atoms moving out of sync with the other two hydrogens, thus
described as asymmetric stretching and asymmetric  bending modes. The [N-2,2]
modes need to have at least four atoms in addition to the center
atom of the molecule. Methane, $CH_4$ has a purely angular mode in an 
$E$ mode of this molecule in which the bonds to the hydrogen atoms show bending
motions that alternately open and close  interparticle
angles.  These motions can be viewed at the following links:
www.chemtube3d.com/vibrationsnh3/ and www.chemtube3d.com/vibrationsch4/

As $N$ increases, these motions evolve in several ways.
There are in fact three ways that
an increase in $N$ can affect the character of the normal modes.

\smallskip

A) As noted above in Section III a.- e., 
the analytic forms of the particle motion  have explicit $N$ or
$\xi$ ($1 \leq \xi \leq N-1$) or $i,j$ dependence
($1 \leq i \leq j-2$, $4 \leq j \leq N$)
(See Eqs.~(\ref{eq:rN}, \ref{eq:SgammaN} 
- \ref{eq:nm2g})
that can affect the character of 
the motion of particles contributing to a particular symmetry
coordinate and thus the normal coordinates. 
 As more particles
are added to the system, the behavior of these larger systems can become
qualitatively quite different from few particle systems with
low values of $N$, e.g. $3 \leq N \leq 6$. 

\smallskip

B) Second, the amount of mixing of the radial and angular 
symmetry coordinates in the
$[N]$ and $[N-1,1]$ sectors (i.e. the values of 
$\cos{\theta^{[N]}_\pm},\, \sin{\theta^{[N]}_\pm}, \, 
\cos{\theta^{[N-1,1]}_{\pm}}$ and
$\sin{\theta^{[N-1,1]}_{\pm}}$ in Eq.~(\ref{eq:mixing})) 
can evolve as $N$ increases.

\smallskip

C) And finally, the frequency of vibration of the normal modes can change as
 a function of $N$.

\smallskip

Although general observations can be made, all three of these
 effects ultimately depend on the particular Hamiltonian of the system. As a
specific example, I will
 look at these effects using a recently studied Hamiltonian for a system of
identical fermions in the unitary regime. 

\subsection{Explicit $N$ dependence}
\label{subsec:general}

In the analytic expressions for the particle motions contributing to the
symmetry coordinates, 
 Eqs.~(\ref{eq:rN}, \ref{eq:SgammaN} 
- \ref{eq:nm2g}),
there is some explicit
$N$ dependence that affects the behavior as $N$ increases.  

\medskip

\subparagraph{The [N] sector.} For the 
symmetric stretch motion in the $[N]$
sector, the character of this breathing motion 
remains the same as $N$ increases with the radial displacements 
decreasing as $N$ increases. The particles move along their 
individual radii out and then in toward the center of the trap, keeping their 
interparticle angles constant as they oscillate about the zeroth order
configuration.

For the angular motion in the $[N]$ sector, the symmetric bend motion
of the interparticle angles for small $N$, (cf. the $A_1$ mode of
ammonia) evolves into a center of mass
motion as $N$ increases. The particles undergo identical angular displacements 
that decrease in size as $N$ increases
 resulting in  motion where the whole ensemble "jiggles" in
response to an excitation of the center of mass mode. 
These jiggles are caused by the particles moving
 past the trap center and then back keeping their radii constant while
 changing their interparticle angles compared to the 
zeroth order configuration. 
As expected, the frequency of this mode, as shown in Section~\ref{sec:Nfreq},
cleanly separates out for all values of $N$ at exactly twice the trap frequency
(The atoms move past the trap center twice in one cycle.), reflecting
the fact the center of mass motion is independent of particle
interactions.

\medskip

For the $[N-1,1]$ and $[N-2,2]$ sectors, the analytic
forms of the 
$(r^{[N-1, \hspace{1ex} 1]}_\xi)_i, (\gamma^{[N-1, \hspace{1ex}1]}_{\xi})_{ij}$ and
$(\gamma^{[N-2, \hspace{1ex}
2]}_{ij})_{mn}$
contain
parameters associated with $N$ that change the character
of the motion as $N$ changes.

\medskip

\subparagraph{The [N-1,1] sector.} As discussed above, in the $[N-1,1]$ sector, the appearance of $\xi$
in the last terms of Eqs.~(\ref{eq:rNm1inr}) and (\ref{eq:gNm1ing})
weights the response of the $\xi +1$st particle. For small values of $N$ and
thus small $\xi$ ($1 \leq \xi \leq N-1$), the motion of this last 
$\xi + 1$st particle, which has opposite direction to the first $\xi$ particles,
appears as a simple asymmetric stretch or bend.  As $N$ becomes large, and
thus $\xi$ can also approach large values, this motion acquires such a large
displacement compared to the remaining particles 
that it is more appropriately  characterized as a 
particle-hole excitation due to
single particle radial or angular motion away from the other
particles ($\xi$ in number) participating in this collective motion. This change
in character from asymmetric stretch (or bend) to single particle 
radial (or angular) excitation
happens quite quickly as $N$ and $\xi$ increase
 with e.g. $N = 10$ and $\xi=N-1=9$ showing obvious
single particle behavior as the ${\xi +1}^{st} =10^{th}$ particle moves with a 
radial displacement
nine times larger than the other $\xi$ particles and in the opposite direction.
Note that the displacements of all $\xi+1$ particles sum to zero 
(1 (particle) $\times$ a displacement of $\xi $ = $\xi$ particles $\times$
 a displacement of 1) reflecting the fact that this motion is a rearrangement
of the particles within the ensemble creating a hole, 
not the loss of a particle due to
radial or angular motion.
\medskip 

\subparagraph{The [N-2,2] sector.} In the $[N-2,2]$ sector which has totally 
angular behavior, a similar
evolution of character is observed as $N$ increases. 
Consider the symmetry coordinate 
${[\bm{S}_{\overline{\bm{\gamma}}'}^{[N-2,2]}]_{ij}}$ where 
($1 \leq i \leq j-2$, $4 \leq j \leq N$), and let $i$ and $j$ assume their 
highest values, $i=N-2$ and $j=N$ so all $N$ particles will be involved in
the motion of this symmetry coordinate 
$[S_{\overline{\gamma}'}^{[N-2,2]}]_{N-2,N}$.
Examining each of the $N(N-1)/2$ angular 
displacements, $(\gamma_{ij}^{\prime [N-2,2]})_{mn}$, for particles $m,n$
 ($1 \leq m < n \leq N$) 
that contribute to the
corresponding 
total angle cosines,
$(\gamma_{ij}^{[N-2,2]})_{mn}=\gamma_{\infty}+\delta^{\frac{1}{2}}
(\gamma_{ij}^{\prime [N-2,2]})_{mn}$,
 it is clear from Eq.~\ref{eq:nm2g} that
there are 3 different magnitudes of $\gamma'$ 
appearing for three types of interparticle
angles.  The displacement $(\gamma_{ij}^{\prime [N-2,2]})_{mn}$ 
is the first-order
correction in $\delta = 1/D$ to the total angle cosine allowing a
determination of the value of the interparticle angle,
$(\gamma_{ij}^{[N-2,2]})_{mn} = \cos \theta_{mn}$. Then 
$\theta_{mn} - \theta_{\infty}$ gives the displacement.  To determine
$\theta_{\infty} = \arccos \gamma_{\infty}$ and thus the displacement, a
specific Hamiltonian must be chosen. 

\smallskip

\paragraph{The dominant interparticle angle.}  When the $m^{th}$ and
$n^{th}$ particles assume the values,
$m=i+1=N-1$ and $n=j=N$, the weighting factors 
$i\delta_{i+1,m}$ and $(j-3) \delta_{jn}$ in Eq.~\ref{eq:nm2g} 
 multiply together producing
a large positive  
factor of $i(j-3)=(N-2)(N-3)$ in the displacement contribution to the
angle cosines, 
$(\gamma_{ij}^{\prime [N-2,2]})_{mn}
=(\gamma_{N-2.N}^{\prime [N-2,2]})_{N-1,N}$.

\smallskip

\paragraph{Nearest neighbor interparticle angles.}  For 
$(\gamma_{ij}^{\prime [N-2,2]})_{mn}$ which have either 
$m=i+1=N-1$ $\bf{or}$ $n=j=N$, (but not both) 
one or the other
weighting factor in Eq.~\ref{eq:nm2g} contributes resulting in a negative 
factor  of $-(N-3)$ in the expression for the displacement.
These $(\gamma_{ij}^{\prime [N-2,2]})_{mn}$ are associated with 
angles $\theta_{mn}$ that are
nearest neighbor interparticle angles 
with the dominant angle $\theta_{N-1,N}$ discussed in part $\it{a.}$
  above.
Thus these interparticle angles are responding to the motion of this 
dominant angle. If the dominant angle is opening, these neighboring angles
are compressing and vice versa.
These angles are:
$\gamma'_{1,N}, \gamma'_{2,N}, \gamma'_{3,N}, \gamma'_{4,N}, .....
\gamma'_{N-2,N}$ and 
$\gamma'_{1,N-1}, \gamma'_{2,N-1}, \gamma'_{3,N-1}, \gamma'_{4,N-1}, .....
\gamma'_{N-2,N-1}$ numbering $2(N-2)$. 
 (I have dropped the
indices $i,j$ referencing the particular symmetry coordinate.)

\smallskip
 
\paragraph{Third type of interparticle angle.}  This leaves $N(N-1)/2-2(N-2)-1$ interparticle angles which have a third type
of displacement for $(\gamma_{ij}^{\prime [N-2,2]})_{mn}$ 
which includes a small factor of $+2$ since neither
weighting factor in Eq.~\ref{eq:nm2g} contributes. 

All the displacements for the particles $m,n$ contributing to the motion of
the $i,j^{th}$ symmetry coordinate include an identical factor of
$[S_{\overline{\gamma}'}^{[N-2,2]}]_{i,j}$ as well as a normalization factor of
$\frac{1}{\sqrt{i(i+1)(j-3)(j-2)}}$.


Note also that all the displacements sum to zero: 
$1 \times (N-2)(N-3)-2(N-2)\times (N-3) +
(N(N-1)/2 -2(N-2)-1)\times 2 \equiv 0$, reflecting the fact that the 
particles
are simply rearranging their positions within the confined angular 
space they occupy.

For very low values of $N$, these expressions yield behavior that is 
qualitatively analogous to the normal mode behavior seen in 
few-body molecular systems like ammonis or methane.
For example, for $N = 4$, there is one dominant interparticle angle, four
nearest neighbor interparticle angles, $2(N-2) = 4$, and a single 
interparticle angle of the third type, $N(N-1)/2-2(N-2)-1=1$.  The
displacements of the dominant angle and the single angle of the third type
are equal for this lowest value of $N$ (the factor $(N-2)(N-3)=2$ for $N=4$)
while the four nearest neighbor angles have a smaller 
displacement (with a factor of $N-3=1$) similar to the behavior of an E mode 
of methane.

As $N$ increases, the relative numbers of the three different 
interparticle angles  
change quickly with the third type of interparticle angle, 
which is not a nearest neighbor of the dominant angle and thus
 has a small response, quickly becoming the overwhelmingly largest
number of interparticle angles.
The first type always has a single dominant angle with
the largest correction to the maximally symmetric zeroth-order configuration.
The second type, which has a noticeable response to the 
opening or closing of the dominant angle, has $2(N-2)$ angles, a number that
increases linearly with $N$, while
the third type of angle which has a negligible response for $N \gg 1$
has $N(N-1)/2-2(N-2)-1$ interparticle 
angles which increases as $N^2/2$, quickly becoming the greatest part of
an ensemble of $N$ particles that is undergoing this collective motion. 
For example, for $N=10$ with $N(N-1)/2=45$ interparticle angles, there is 
a single dominant angle, 16 angles that are nearest neighbors and 28
angles with a very small response in the third group. For $N=100$, 
there are 4950 interparticle angles: one dominant
angle, 196 nearest neighbor angles that show a noticeable response
  and  4753 angles that have a very small
response.  

Thus a picture emerges as $N$ increases of 
compressional or phonon behavior for the motion in this $[N-2,2]$ sector.
These $[N-2,2]$ modes involve oscillations in the angles 
that push the atoms together 
and pull them apart with no change in the
particles' radial positions. 
This type of motion is best characterized as a compressional stationary wave
i.e. a phonon oscillation.
This is consistent with the very low frequency of this mode compared to
the frequencies of the other four types of 
normal coordinates as will be shown in
Section~\ref{sec:Nfreq} and the large zero-point energy as seen in
Eq.~(\ref{eq:E1}). 

\medskip

\subparagraph{Examples.} The three types of interparticle angles have 
{\it displacement} values that also
depend on
the value of $N$, evolving from displacements that are roughly 
comparable for all three types
of angles for low values of $N$, e.g. $4 \leq N \leq 6$ 
to displacements that are 
quite different in magnitude differing by factors $\sim N$ and $\sim N^2$
as $N$ increases. Using the ratios of the different factors discussed above
in the expressions for the angular displacements, 
the dominant angle has an angular
displacement value of $(N-2)(N-3)$ that is a factor of $N-2$ times 
larger than the angular 
displacements of $-(N-3)$ of the
nearest neighbor angles  
and is a factor of $(N-2)(N-3)/2$ times larger than the 
angular displacements of 2 for the third type of interparticle angle. As 
explicit examples,
I will look at the ratios of the three types of displacements
 for two values of $N$: $N=4$ and $N=10$
and  will again consider the motion of individual particles participating in the
collective motion of the symmetry coordinate 
$[S_{\overline{\gamma}'}^{[N-2,2]}]_{ij}$ which has the highest values of 
$i$ and $j$  ($i=N-2, j=N$) and thus involves the motion of all $N$
particles.
For $N=4$, the symmetry coordinate is expected to
have behavior similar to an E mode of methane; and  for $N=10$,  it
will be seen that the behavior of this symmetry coordinate in the
$[N-2,2]$ sector has already 
evolved into
 compressional behavior.

For the lowest value of $N=4$, the displacement of the dominant angle is twice
(the factor $N-2 = 2$) the displacement of the nearest neighbor angles 
and equal to the displacement (with factor $(N-2)(N-3)/2 = 1$)
 of the single angle in the third type.
Thus, the dominant angle and the single angle of the third type have a
contribution to the angle cosine that is twice as large as the four nearest
neighbor angles.
The interparticle angle $\theta_{12}$
associated with particles 1 and 2 opens and closes by the same amount and
in sync with the dominant
angle $\theta_{34}$  between particles 3 and 4.
The nearest neighbor angles, $\theta_{13}, \theta_{14}, \theta_{23}, \theta_{24}$,
open (and then close) by an amount that is roughly half as large
 in response to the closing (opening)
 of the dominant angle. (Since the value of $\gamma_{\infty}$ is typically
close to zero which is the value for mean field interactions,
 the angular displacements to the angle cosines 
roughly give the actual angles of these motions.)

Thus, the four particles involved in the collective motion of the
symmetry coordinate ${[\bm{S}}_{\overline{\bm{\gamma}}'}^{[N-2, \hspace{1ex} 2]}]_{24}$
perform a simultaneous opening and closing of two different interparticle
angles, $\theta_{34}$ and $\theta_{12}$ similar to an $E$ mode of methane
with the neighboring angles making smaller adjustments.

A similar analysis for the case of $N=10$ shows clearly that the motion of the
particles participating in this symmetry coordinate has evolved from the
methane picture of two interparticle angles opening and closing in sync,
to behavior that looks much more like a compressional wave.
For $N=10$ and letting $i$ and $j$ assume their largest values, $i=8, j=10$, 
I analyze
the behavior of the particles participating in the symmetry coordinate
${[\bm{S}}_{\overline{\bm{\gamma}}'}^{[N-2, \hspace{1ex} 2]}]_{8,10}$.
The dominant
interparticle angle $\theta_{9,10}$ closes (and then opens) with an angular
displacement that is eight times (the factor $N-2 =8$)
 the angular displacement of the 16 nearest
neighbor angles and 28 times (the factor $(N-2)(N-3)/2 = 28$) the angular
displacement of the
28 interparticle angles of the third type. Thus both the nearest neighbor
angles and especially the third type of interparticle
angle which has become the largest group have already begun to have
a very small response compared to the change in the dominant angle.

This trend is expected to continue as $N$ increases and the number of 
interparticle angles in the third type becomes very large. (For $N=100$,
the dominant angle closes by an angular displacement that is 98 times
(the factor $N-2 =98$) larger than the 196 nearest neighbor angles which by
open a very small amount,  
while the 4753 interparticle angles of
the third type adjust by an even smaller amount, 4753 times
(the factor $(N-2)(N-3)/2 = 98\times 97/2 = 4753$) smaller than the dominant 
angle,
more than three orders of magnitude smaller, clearly a negligible response.

\medskip

\subsection{Mixing coefficients as a function of N}

For the mixing coefficients that determine the radial/angular mixing in the
normal modes for the $[N]$ and $[N-1,1]$ sectors, very few general comments
can be made without specifying a particular Hamiltonian.
The mixing coefficients as defined in Eq.~(\ref{eq:tanthetaalphapm}) 
have a complicated $N$ dependence that originates in the Hamiltonian
terms at first order. All the terms in the Hamiltonian have
explicit $N$ dependence which affects the mixing of the radial and angular
modes of the $[N]$ and $[N-1,1]$ sectors. In particular, the type of confining
potential and the particular interparticle interaction chosen will affect
the $N$ dependence of the normal modes' behavior.  
(Of course, the type of confining potential and interparticle interaction
potential have effects on the character of the normal modes through the 
mixing coefficients (as well as the frequencies) apart from the $N$ dependence
that I am studying in this paper.)

In the case of the recently studied system
 of identical fermions in the unitary regime which has been intensely
studied in the laboratory, 
the mixing coefficients for the 
$[N]$ sector have the
following form:

\begin{widetext}
\begin{eqnarray}
\mbox{cos}\theta^{[N]}_+ &=& \frac{\sqrt{2} \sqrt{N-1}(c+(N/2-1)d)}
{\sqrt{2(N-1)(c+(N/2-1)d)^2
+(-a-(N-1)b+{\lambda}_{[N]}^+)^2}}\label{eq:cos0p} \\
\mbox{sin}\theta^{[N]}_+ &=& \frac{-a-(N-1)b+{\lambda}_{[N]}^+}
{\sqrt{2(N-1)(c+(N/2-1)d)^2+(-a-(N-1)b
+{\lambda}_{[N]}^+)^2}} \label{eq:sins0p}\\
\mbox{cos}\theta^{[N]}_- &=& \frac{\sqrt{2} \sqrt{N-1}(c+(N/2-1)d)}
{\sqrt{2(N-1)(c+(N/2-1)d)^2
+(-a-(N-1)b+{\lambda}_{[N]}^-)^2}}\label{eq:cos0m} \\
\mbox{sin}\theta^{[N]}_- &=& \frac{-a-(N-1)b+{\lambda}_{[N]}^-}
{\sqrt{2(N-1)(c+(N/2-1)d)^2+(-a-(N-1)b
+{\lambda}_{[N]}^-)^2}} \label{eq:sin0m}
\end{eqnarray}
\end{widetext}

\noindent where ${\lambda}_{[N]}^\pm$ is given by Eq.~\ref{eq:lambda12pm}.
The quantities $a,b,c,d, \mbox{ and } {\lambda}_{[N]}^{\pm}$, are defined
in Eq.(42) in Ref.~\cite{FGpaper} in terms of the $F$ and $G$ elements
and have explicit $N$ dependence
as well as $N$ dependence from the $F$ and $G$ elements of Eq.~\ref{eq:Gham}    
from the specific Hamiltonian.  These
$F$ and $G$ elements are defined in Ref.~\cite{FGpaper} 
for three different
Hamiltonians in Eqs. (75, 76, 100, 101, 119, 120) and exhibit 
explicit $N$ dependence 
that originates in the Hamiltonian
terms at first order. Thus there are three layers of analytic expressions
that can bring in $N$ dependence: the expressions for $\cos\theta^{[N]}_{\pm}$
and $\sin\theta^{[N]}_{\pm}$ in Eqs.~(\ref{eq:cos0p}-\ref{eq:sin0m}) above, 
the expressions for $a,b,c,d, \mbox{ and } \lambda_{[N]}^{\pm}$
and the expressions for the $F$ and $G$ elements for a specific Hamiltonian.

\begin{figure}
\includegraphics[scale=0.9]{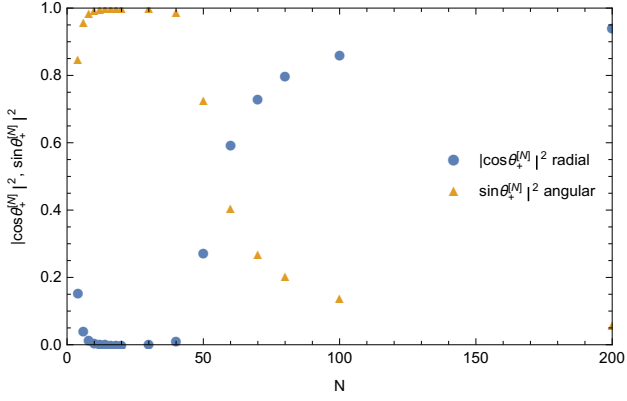}
\renewcommand{\baselinestretch}{0.8}
\caption{The square of the mixing coefficients $|\cos\theta^{[N]}_+|^2$ 
and $|\sin\theta^{[N]}_+|^2$ for the normal mode $q^{\prime [N]}_+$
 as a function of N.}
\label{fig:one}
\end{figure}

\begin{figure}
\includegraphics[scale=0.9]{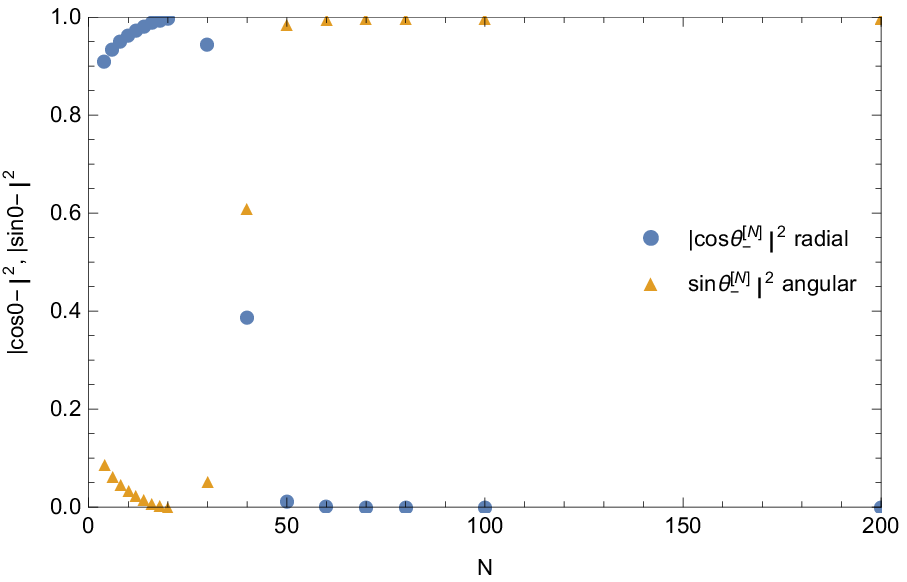}
\renewcommand{\baselinestretch}{0.8}
\caption{The square of the mixing coefficients $|\cos\theta^{[N]}_-|^2$ 
and $|\sin\theta^{[N]}_-|^2$ for the normal mode $q^{\prime [N]}_-$
 as a function of N.}
\label{fig:two}
\end{figure}

\begin{figure}
\includegraphics[scale=0.9]{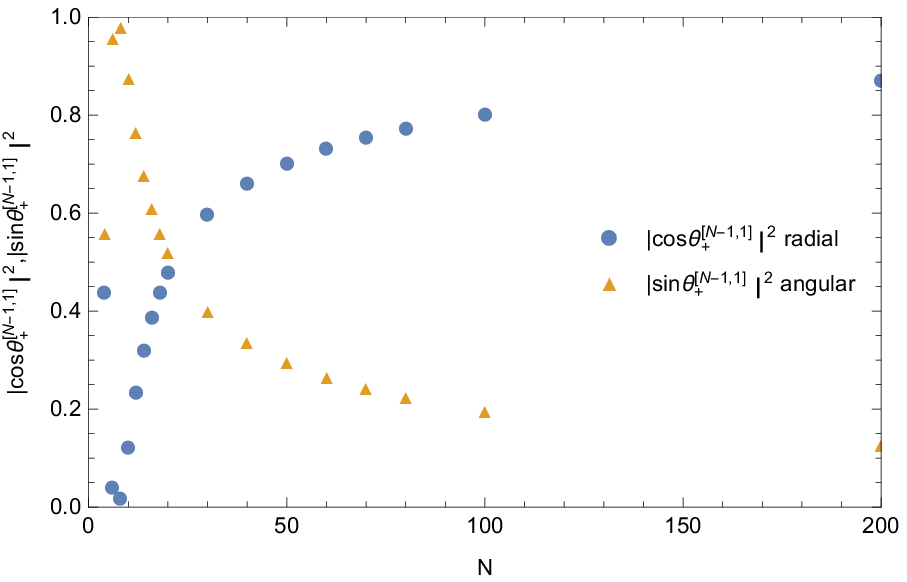}
\renewcommand{\baselinestretch}{0.8}
\caption{The square of the mixing coefficients $|\cos\theta^{[N-1,1]}_+|^2$ 
and $|\sin\theta^{[N-1,1]}_+|^2$ for the normal mode $q^{\prime [N-1,1]}_+$
 as a function of N.}
\label{fig:three}
\end{figure}

\begin{figure}
\includegraphics[scale=0.9]{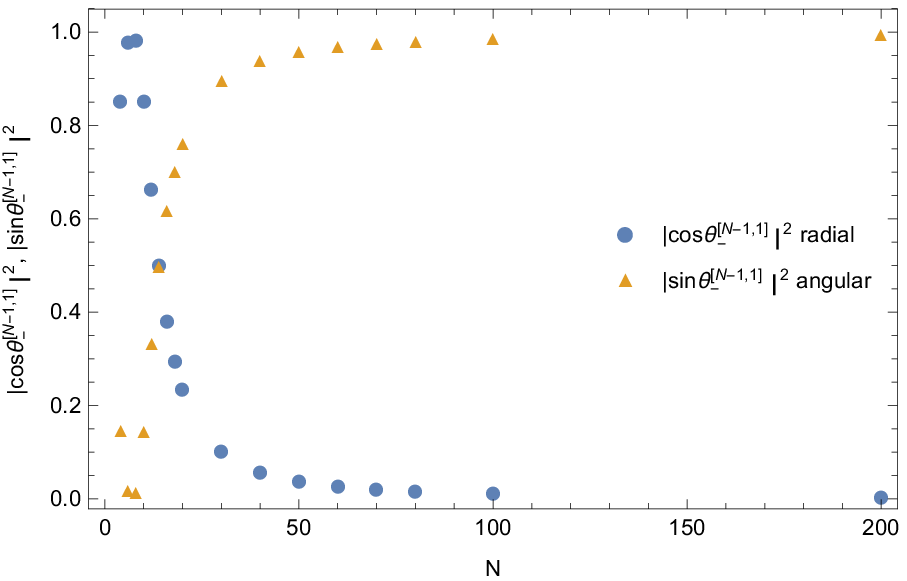}
\renewcommand{\baselinestretch}{0.8}
\caption{The square of the mixing coefficients $|\cos\theta^{[N-1,1]}_-|^2$ 
and $|\sin\theta^{[N-1,1]}_-|^2$ for the normal mode $q^{\prime [N-1,1]}_-$
 as a function of N.}
\label{fig:four}
\end{figure}

I show the resulting behavior of the mixing coefficients as a function of
$N$ for a system of identical fermions in the unitary regime in 
Figs.~(\ref{fig:one}-\ref{fig:four}).
In Fig.~\ref{fig:one}, I have plotted the square of the mixing coefficients, 
${|\cos\theta^{[N]}_+|^2}$ and ${|\sin\theta^{[N]}_+|^2}$ for
$q^{\prime [N]}_+$:

\begin{equation} 
{\bm{q}'}_+^{[N]} = c_+^{[N]} \left(
\cos{\theta^{[N]}_+} \, [{\bm{S}}_{\bar{\bm{r}}'}^{[N]}]
\, + \, \sin{\theta^{[N]}_+} \,
[{\bm{S}}_{\overline{\bm{\gamma}}'}^{[N]}] \right) \nonumber
\end{equation}

\noindent as a function of $N$. The square of
these coefficients gives the probability associated with each symmetry
coordinate, $[{\bm{S}}_{\bar{\bm{r}}'}^{[N]}]$ or
$[{\bm{S}}_{\overline{\bm{\gamma}}'}^{[N]}] $,
 in the expression for the 
normal mode, $q^{\prime [N]}_{+}$. 
The plot shows that the character of the normal mode 
 $q^{\prime [N]}_{+}$ is almost purely angular for $N \lesssim 30$ after
which there is a gradual crossing of character between $N=30$
and
$N=200$ when the normal mode has mixed radial and angular character. 
For $N \geq 200$ the character is $> 90\%$ radial and for 
$N \gg 1$, the character becomes almost purely radial.

The other normal mode in the $[N]$ sector, $q^{\prime [N]}_-$ has complementary
behavior, starting out totally radial and switching to totally angular
as shown in Fig.~\ref{fig:two}. In this case,, the crossover is quite sharp,
starting again around $N \backsim 30$, but finishing the crossover by
$N=50$.
So this normal mode, $q^{\prime [N]}$, is purely radial for low $N$ and
purely angular for very large values of $N$.
It has mixed radial/angular character for a very small range of $N$.

A bit of inspection reveals that this behavior is being dictated 
to a large extent by the explicit $N$
dependence in the expressions
for $\cos\theta^{[N]}_{\pm}$ and $\sin\theta^{[N]}_{\pm}$ 
( Eqs.~(\ref{eq:cos0p}-\ref{eq:sin0m}) above) which have alternating
limits of 0 or 1 as $N\rightarrow 1$ or $N\rightarrow \infty$.
The position and shape of the crossover is influenced by the other 
sources of $N$ dependence that originate in the specific Hamiltonian.

The mixing coefficients for the 
$[N-1,1]$ sector have the
following form:

\begin{widetext}
\begin{eqnarray}
\mbox{cos}\theta^{[N-1,1]}_+ &=&  \frac{\sqrt{N-2}(c-d)}
{\sqrt{(N-2)(c-d)^2 + (-a+b+{\lambda}_{[N-1,1]}^+)^2}}\label{eq:cos1p} \\
\mbox{sin}\theta^{[N-1,1]}_+ &=& \frac{-a+b+{\lambda}_{[N-1,1]}^+}
{\sqrt{(N-2)(c-d)^2 + (-a+b+{\lambda}_{[N-1,1]}^+)^2}}\label{eq:sin1p} \\
\mbox{cos}\theta^{[N-1,1]}_- &=& \frac{\sqrt{N-2}(c-d)}
{\sqrt{(N-2)(c-d)^2 + (-a+b+{\lambda}_{[N-1,1]}^-)^2}}\label{eq:cos1m} \\
\mbox{sin}\theta^{[N-1,1]}_- &=&  \frac{-a+b+{\lambda}_{[N-1,1]}^-}
{\sqrt{(N-2)(c-d)^2 + (-a+b+{\lambda}_{[N-1,1]}^-)^2}}\label{eq:sin1m} 
\end{eqnarray}
\end{widetext}

\noindent where the quantities $a,b,c,d, \mbox{ and } 
{\lambda}_{[N-1,1]}^{\pm}$, as well as the $F$ and $G$ elements
 are defined as before.

In Fig.~\ref{fig:three}, I have plotted the square of the mixing coefficients 
for $q^{\prime [N-1,1]}_{\xi +}$, again for a Hamiltonian describing a system of 
identical
fermions in the unitary regime.
The plot shows that the character of the normal mode 
 $q^{\prime [N-1,1]}_{\xi +}$ is almost purely angular for $N \lesssim 10$ after
which there is a rather sudden crossing of character 
and then a gradual trend toward  purely radial character.
For $N \gg 1$, the character is almost purely radial.
  The other normal mode in the $[N-1,1]$ sector, $q^{\prime [N-1,1]}_{\xi -}$ 
has complementary
behavior, starting out totally radial and switching to totally angular
as shown in Fig.~\ref{fig:four}. In this case the crossing is quite sharp.

Analogous to the $[N]$ sector, this behavior is being dictated 
to a large extent by the explicit $N$
dependence in the expressions
for $\cos\theta^{[N-1,1]}_{\pm}$ and $\sin\theta^{[N-1,1]}_{\pm}$ 
in  Eqs.~(\ref{eq:cos1p}-\ref{eq:sin1m}) above which have alternating
limits of 0 or 1 as $N\rightarrow 2$ or $N\rightarrow \infty$.
The position and shape of the crossover is influenced by the other 
sources of $N$ dependence that originate in the specific Hamiltonian.

\subsection{Normal mode frequencies as a function of $N$.}
\label{sec:Nfreq}
Analytic expressions for the normal mode frequencies were derived in 
Ref.~\cite{FGpaper} using a
method  outlined
in Appendices B and C of that paper which derives analytic formulas for the 
roots, $\lambda_{\mu}$, of 
the $FG$ secular equation.
The  normal-mode vibrational
frequencies, $\bar{\omega}_{\mu}^2$, are related to the roots $\lambda_{\mu}$ of
${\bf FG}$ by:
\begin{equation}\label{eq:omega_p}
\lambda_{\mu}=\bar{\omega}_{\mu}^2,
\end{equation}
The two  frequencies associated with the $\lambda_{0}$ roots of 
multiplicity one are of the form
\begin{equation}
\bar{\omega}_{{0}^{\pm}}=\sqrt{\eta_0 \pm
\sqrt{{\eta_0}^2-\Delta_0}},
\end{equation}

\noindent where:

\begin{equation}\label{eq:lam0defs}
\begin{split}
\eta_0 &= \frac{1}{2}\Biggl[a-(N-1)b+g+2(N-2)h \\
& \,\,\,\,\,\,\,\,\,\,\,\,\,\,\,\,\,\,\,\,\,\,\,\,\,\,\,\,\,\,\,\,\, +\frac{(N-2)(N-3)}{2}\iota \Biggr] \\ 
\Delta_0 &= (a-(N-1)b)\left[g+2(N-2)h-\frac{(N-2)(N-3)}{2}\iota\right]  \\
&\,\,\,\,\,\,\,\,\,\,\,\,\,\,\,\,\,\,\,\,\,\,\,\,\,\,\,-\frac{N-2}{2}(2c+(N-2)d)(2e+(N-2)f).
\end{split}
\end{equation}

%
For the two $N-1$ multiplicity roots, the frequencies are of the
form
\begin{equation}
\bar{\omega}_{{1}^{\pm}}=\sqrt{\eta_1 \pm
\sqrt{{\eta_1}^2-\Delta_1}},
\end{equation}

\noindent where $\eta_1$ and $\Delta_1$ are given by:
\begin{eqnarray}\label{eq:lam1defs}
\eta_1 &=& \frac{1}{2}\left[a-b+g+(N-4)h-(N-3)\iota\right] \nonumber \\
\Delta_1 &=&(N-2)(c-d)(e-f)+(a-b) \nonumber \\
   &&\times\left[g+(N-4)h-(N-3)\iota\right].
\end{eqnarray}
The frequency ${\bar{\omega}}_2$, associated with the root $\lambda_2$ of
multiplicity $N(N-3)/2$ is given by:
\begin{equation}
\bar{\omega}_2=\sqrt{g-2h+\iota}.
\end{equation}
The quantities $a,b,c,d,e,f,g,h,i$ are defined as before 
in Eq. (42) in Ref~\cite{FGpaper}
in terms of the $F$ and $G$ elements and have explicit $N$ dependence as well
as $N$ dependence from the $F$ and $G$ elements from a particular Hamiltonian.

Thus the analytic expressions for the frequencies 
(like the mixing coefficients) have three layers of 
$N$ dependence: explicit $N$ dependence in the formulas for $\eta_0$,
$\Delta_0$, $\eta_1$ and $\Delta_1$ in Eqs~(\ref{eq:lam0defs}) and
(\ref{eq:lam1defs}); explicit $N$ dependence in the formulas for
the quantities $a,b,c,d,e,f,g,h,i$, and finally the $N$ dependence 
in the $F$ and $G$
elements from
a specific Hamiltonian in these formulas.

\begin{figure}
\includegraphics[scale=0.9]{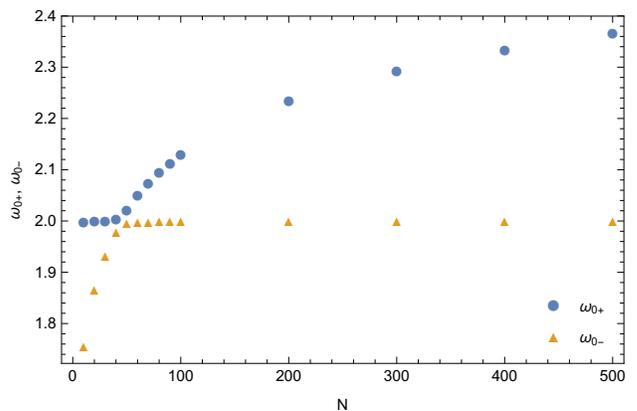}
\renewcommand{\baselinestretch}{0.8}
\caption{The frequencies $\bar{\omega}_{{0}^{\pm}}$ in units of the trap
frequency
for the normal modes $q^{\prime [N]}_{\pm}$
 as a function of N.}
\label{fig:five}
\end{figure}

\begin{figure}
\includegraphics[scale=0.9]{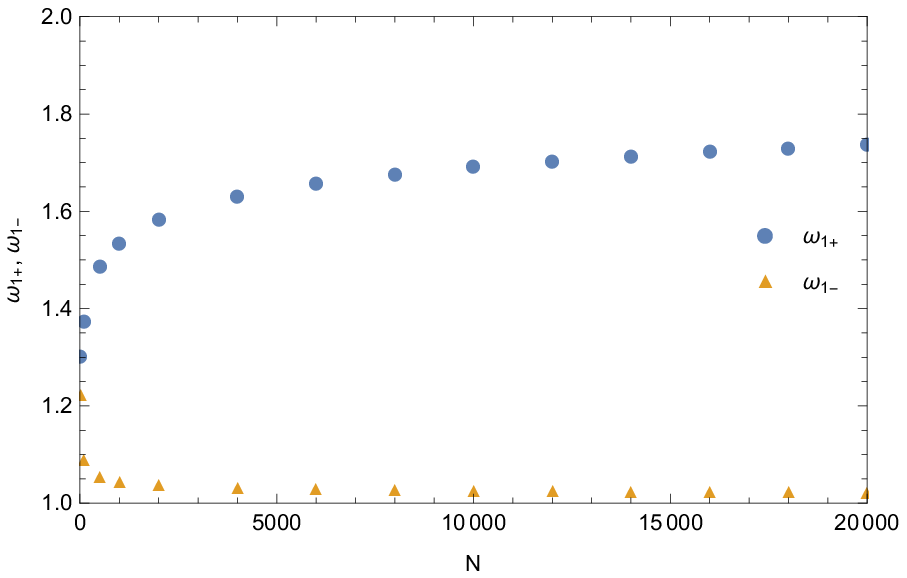}
\renewcommand{\baselinestretch}{0.8}
\caption{The frequencies $\bar{\omega}_{{1}^{\pm}}$ in units of the trap
frequency
 for the normal modes $q^{\prime [N-1,1]}_{\pm}$
 as a function of N.}
\label{fig:six}
\end{figure}

\begin{figure}
\includegraphics[scale=0.7]{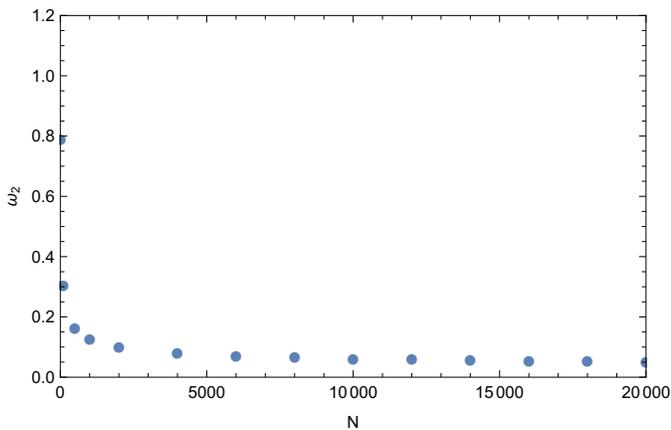}
\renewcommand{\baselinestretch}{0.8}
\caption{The frequency $\bar{\omega}_{2}$ in units of the trap
frequency
 for the normal mode $q^{\prime [N-2,2]}$
 as a function of N.}
\label{fig:seven}
\end{figure}

In  Figs.~(\ref{fig:five})-(\ref{fig:seven}), I show the $N$ dependence of the
frequencies for the five types of normal modes for a Hamiltonian 
of an ensemble of 
identical
fermions in the unitary regime.

In Fig.~(\ref{fig:five}), the frequencies  $\bar{\omega}_{{0}^+}$
and $\bar{\omega}_{{0}^-}$ in the $[N]$ sector show a clear avoided crossing
as the characters of the normal modes $q^{\prime [N]}_+$ and $q^{\prime [N]}_-$
change  from angular to radial for $q^{\prime [N]}_+$
and radial to angular for $q^{\prime [N]}_-$. The radial behavior is
associated with a frequency that  starts below the center of mass
frequency for low $N$ and then rises above the center of mass frequency
as $N$ increases. Note that the 
angular behavior is  associated with a frequency that is exactly
twice the trap frequency for all values of $N$ revealing the
separation of a center of mass coordinate. As analyzed in 
Section~\ref{sec:symmetrymotions}, this angular motion in the $[N]$ sector
looks like a symmetric bending motion for small values of $N$ as seen in the
$A_1$ mode of ammonia, but evolves into a rigid center of mass movement of 
the whole ensemble with the radial interparticle distances remaining rigidly 
constant
as the entire ensemble moves relative to the center of the trap
creating small displacements for the interparticle angles.

In Fig.~(\ref{fig:six}), the frequencies  $\bar{\omega}_{{1}^+}$
and $\bar{\omega}_{{1}^-}$ in the $[N-1,1]$ sector show behavior that starts
out with both frequencies $\sim 1.3$ times the trap frequency 
for low values of $N$.  As $N$ increases these two
frequencies rapidly separate, one, $\bar{\omega}_{{1}^-}$, associated with
angular behavior, going to the trap frequency and the other,
 $\bar{\omega}_{{1}^+}$, describing radial behavior, increasing slowly.
 In both these sectors, $[N]$ and
$[N-1,1]$, the radial frequency is higher than the corresponding angular
frequency which is expected and is also seen for the small molecules like 
ammonia and methane
that offer good molecular equivalents.

In Fig.~(\ref{fig:seven}), the frequency  $\bar{\omega}_{2}$
of the $[N-2,2]$ sector shows behavior that starts
out for low values of $N$ at values near the trap frequency.  
As $N$ increases this frequency rapidly decreases to extremely small values,
several orders of magnitude smaller than the trap frequency
consistent with the slow oscillations of a phonon mode.

Inspection of the sources of $N$ dependence reveal that 
the frequencies shown in 
Figs.~(\ref{fig:five})-(\ref{fig:seven}), for a  Hamiltoninan of identical
fermions in the unitary regime 
show a complicated dependence
on all three layers of  $N$ dependence.  
This result differs from the behavior of the mixing coefficients for
the same Hamiltonian which showed behavior as a function of $N$ that
was dominated by the explicit $N$ dependence in  
Eqs.~(\ref{eq:cos0p}-\ref{eq:sin0m})
and Eqs.~(\ref{eq:cos1p}-\ref{eq:sin1m}) and was not as sensitive to
the other sources of $N$ from the specific Hamiltonian.

\section{Summary and Final Thoughts} \label{sec:SumConc}


In this study, I have looked in detail at both the macroscopic collective
behavior and the microscopic contributions of individual particles to 
this behavior
for the normal mode solutions to the symmetry-invariant perturbation
first order equation in inverse dimensionality for a
system of confined, interacting identical particles.  
These normal mode solutions were previously 
obtained analytically as a function of $N$ 
and used to obtain accurate results for energies,
frequencies, wave functions, and
density profiles for systems of 
identical bosons \cite{energy,test,toth,laingdensity} and energies,
frequencies\cite{prl}
 and
thermodynamic quantities\cite{emergence}
 for ultracold fermions in the unitary regime.

These solutions have been tested against an exactly solvable
 model problem
of harmonically interacting particles under harmonic confinement\cite{test}. 
Comparing this wave
function to the exact analytic wave function obtained in an 
independent solution, 
exact 
agreement was found (to ten or more digits of accuracy),
 confirming this general theory for a fully interacting 
$N$-body system in three dimensions\cite{test} and verifying the
analytic expressions for this normal mode basis.
We also tested this general, fully interacting 
wave function for bosons of Ref.~\cite{JMPpaper}, exact through first order, 
by deriving
a property, the density profile of the ground state,  
for the same model problem. 
Our density profile
is indistinguishable from the $D=3$ first-order result 
from the independent solution of this fully interacting, N-body
problem\cite{toth}.

These earlier studies verifying the general formalism, did not focus
on the physical character of this symmetry basis used to obtain
the normal modes solutions. As mentioned earlier, our construction
of the symmetry coordinates  was
done systematically as described in Ref.~\cite{JMPpaper} so the symmetry
coordinates which transform irreducibly under $S_N$ have the
simplest functional forms possible. The first symmetry coordinate
involves only two of the particles and each
succeeding symmetry coordinate was chosen to have the next simplest
functional form possible under the requirement that it transforms irreducibly
 under $S_N$ etc. With this choice the
complexity of the motions described by the symmetry coordinates was
kept to a minimum,  building up incrementally as additional particles were
involved in the motion, ensuring that there
was no
disruption of lower $N$ symmetry coordinates. This process was chosen
primarily to simplify the mathematical 
complexity of this basis. Other choices
would have resulted in different mathematical functions that still comprised
a basis for the normal mode solutions in each sector. Note that the symmetry
coordinates depend only on the symmetry structure of the Hamiltonian, not on 
specific details of the interparticle potential, unlike the normal mode 
coordinates which depend on the specific details of the potentials involved.

Our initial studies using these symmetry coordinates were focused 
on ground states of systems of ultracold 
bosons\cite{energy,laingdensity} and later
fermions\cite{prl} The first use of excited states using this many-body
formalism was in a recent paper studying thermodynamic
quantities for identical fermions in the unitary regime. 
Constructing the partition function in this study
required the use of a large number of excited states from the 
normal mode spectrum (which has an infinite number of equally spaced states). 
These states are chosen specifically to comply 
with the enforcement of the
Pauli principle, thus connecting the Pauli principle to 
many-body interaction dynamics through the normal modes. 
The success of this study in obtaining 
thermodynamic
quantities for the energy, entropy and heat capacity that agree quite
well with experimental data has increased the interest in investigating the
physical character of these states since they
offer the possibility of acquiring physical intuition into the dynamics
of the collective motion supported by this unitary regime.
In particular, the
phonon character of the normal modes with the lowest frequency and the
radial (or angular) excitation of a single particle out of this mode,
i.e. a particle-hole excitation, present a
picture of the dynamics that leads to a gapped spectrum
and collective behavior in the form of superfluidity. With this 
motivation, I have investigated closely both the macroscopic behavior of each
of the five types of 
normal modes and the microscopic contributions of each particle
to this collective behavior, studying the evolution of collective motion
as the number of particles increases.

\subparagraph{Summary.} In summary, my analysis has shown a consistent picture of behavior 
evolving smoothly and rapidly from the low
$N$ systems that have good molecular equivalents, as seen in the behavior
of ammonia and methane
to very different character for the collective motion of larger $N$ systems.
A number of observations have been made from this analysis that may prove
useful in understanding the contribution of particle behavior to the
emerging
collective behavior of an ensemble. I list them below:

\smallskip

1) The analytic expressions for the normal modes produce behavior for small
$N$ that is analogous to the known behavior of small molecular systems
such as ammonia and methane whose atoms move under Coulombic confinement.

\smallskip

2) As $N$ increases, the behavior of these same analytic functions 
rapidly changes character, with the exception of the symmetric
stretch/breathing motion (part i) below.

\smallskip

\indent\indent i) In the $[N]$ sector, the breathing motion of the radial $[N]$ mode retains 
this character as $N$ increases with the symmetric radial displacements simply 
decreasing in amplitude.

\smallskip

\indent\indent ii) The angular mode in the $[N]$ sector evolves
 from a symmetric bending
character for small $N$ to a center of mass motion for large $N$. This
change in character occurs for fairly small $N$. For example, for 
$N = 10$, the motion
would be viewed more appropriately 
as a center of mass rigid motion of the $N(N-1)/2 = 45$
interparticle angles making identical small adjustments, rather than 
viewed as a symmetric
bending motion.

\smallskip

\indent\indent iii) and iv) The asymmetric stretch and asymmetric
bending character of the 
normal modes in the 
$[N-1,1]$ sector
for low $N$ evolves smoothly into radial and angular single particle
excitations, i.e. particle-hole excitations. Again, this happens quickly as $N$ increases. By $N=10$,
 the $10^{th}$ particle participating in the motion of the symmetry
coordinate $[{\bm{S}}_{\bar{\bm{r}}'}^{[N-1,\hspace{1ex} 1]}]_{\xi=9}$ or
$[{\bm{S}}_{\overline{\bm{\gamma}}'}^{[N-1, \hspace{1ex} 1]}]_{\xi=9}$
has a  displacement that is nine
times the displacements of the other particles resulting in behavior
that is viewed more appropriately
as a radial or angular excitation. 
 
\smallskip

\indent\indent v) Similarly, the bending modes of the small $N$ systems
in the $[N-2,2]$ sector quickly and smoothly evolve into
  compressional or phonon behavior as $N$ increases.

\smallskip

Interestingly, this change in character from small $N$ to large $N$
is dictated by fairly simple analytic forms that 
create the evolution  in character for the symmetry coordinates.
Inspecting the forms of the microscopic motions of the individual particles
for the different sectors: $[N], [N-1,1],$ and $[N-2,2]$ 
(See Eqs.~(\ref{eq:rN}, \ref{eq:SgammaN}, \ref{eq:rNm1inr}- \ref{eq:nm2g}),
the relevant $N$ dependence in the 
$[N]$ sector is just in a normalization
factor creating a decrease in amplitude as $N$ increases for these  
symmetric motions. However in the $[N-1,1]$ and $[N-2,2]$ sectors, the
functional $N$ dependence is more involved.  Ignoring the leading
common factors including a normalization factor, the $N$ dependence is
determined by an intricate balancing of Kronecker delta functions and Heaviside
functions that give zero or unity depending on the value of their indices
which involve integers referring to specific particles. This intricate
accounting
of $1's$ and $0's$ determines the motion of the $N$ particles of this normal
mode for both small $N$ and large $N$. 
Not surprisingly, the character depends on knowing the interplay of
all the individual particles one by one, whose contributions are
kept track of perfectly by the Kronecker delta
and Heaviside functions.

3) The behavior of the normal modes which have
 a mixture of the radial and angular symmetry coordinates
in the $[N]$ and $[N-1,1]$ sectors  was investigated for a particular
Hamiltonian of current interest, that of an ensemble of identical confined
fermions in the unitary regime. This behavior 
is seen to transition from totally radial, i.e. a pure radial symmetry
coordinate to totally angular behavior, i.e. a pure angular symmetry coordinate
(or vice versa) as $N$ changes.
Thus for very small $N$, ($N \le 20$ for the $[N]$ sector and $N \le 10$
for the $[N-1,1]$ sector)
 or for very large $N$, $N \gg 1$, the normal modes adopt the 
character of a pure
symmetry coordinate displaying either totally radial or totally angular
behavior.  For some intermediate values of $N$ there is a region where the
normal coordinates show significant mixing of the radial and angular
symmetry coordinates making the behavior more difficult to characterize.  
In some cases, when the crossing is very sharp,
this region is quite small, while other cases show a broader evolution of
character from radial to angular or angular to radial. For large $N$, the
normal modes evolve into purely radial or purely angular behavior in the
case of identical, confined fermions in the unitary regime.
This means that the analytically derived symmetry coordinates are 
eigenfunctions of this first order perturbation equation.
When the Hamiltonian is transformed
into block diagonal form by the symmetry coordinates, the off diagonal elements
are negligible in these regimes. 

This result has implications for the 
stability of collective behavior in this regime since the symmetric
coordinates are eigenfunctions of an approximate underlying Hamiltonian
and thus have some degree of stabiltiy unless the sytem is perturbed. e.g.
by an increase in temperature etc.

Although the construction of the symmetry coordinates was chosen to 
minimize the mathematical complexity and is not a unique basis of 
coordinates, the  symmetry coordinates clearly contain 
information about the dynamics of this many-body problem of identical
particles. Regardless of the
strategy of their construction, they are, by definition and by construction,
coordinates that transform under the irreducible representations of the
symmetric group of $N$ identical objects so the Hamiltonian of this first
order equation which is invariant under the $N!$ symmetry operations 
of the symmetric group is transformed to block diagonal form when expressed
in this basis. In this case, the blocks are small, $2 \times 2$, in the $[N]$
and $[N-1,1]$ sectors that have both radial and angular representations and
$1 \times 1$ blocks, i.e. diagonal for the $[N-2,2]$ sector.

This Hamiltonian term in the first order perturbation equation contains 
beyond-mean-field effects. Thus the normal coordinates whose frequencies
and mixing coefficients depend on the interparticle interactions are, in fact, 
beyond-mean-field ${\it analytic}$ solutions to a many-body Hamiltonian 
through first order.

\smallskip

4) Except for the center of mass frequency which separates out for all values
of $N$ at twice the trap frequency, the frequencies of oscillation of the 
normal modes  also evolve as $N$ increases.
For the case studied of fermions in the unitary regime, the radial frequencies 
 increase in both the $[N]$ and $[N-1,1]$
sectors while the angular frequencies trend toward the trap frequency 
in the case of the $[N-1,1]$ sector and very small values
 in the case of the phonon mode in the $[N-2,2]$
sector. (See Figs. 5 -7)
The frequency of the $[N-2,2]$ modes which starts out for low $N$ as a
bending mode with a frequency near the trap frequency quickly decreases
as the motion evolves into phonon compressional 
behavior, going to extremely small values
for this low energy mode, several orders of magnitude smaller than the trap
frequency. (See Fig. 7.)

\smallskip

5) The normal coordinates provide a basis not just for the ground state, but for
the spectrum of excited states for $L=0$ and for higher order corrections
in the perturbation expansion. I analyzed just one of the $N-1$ 
degenerate symmetry coordinates in each of the $[N-1,1]$ sectors that have
a frequency of $\bar{\omega}_1$ 
and just one of the $N(N-3)/2$
degenerate symmetry coordinates in the $[N-2,2]$ sector that have a 
frequency
of $\bar{\omega}_2$.  In these cases, the symmetry coordinate was chosen
to have the highest indice(s) possible: $\xi = N-1$ for 
$[{\bm{S}}_{\bar{\bm{r}}'}^{[N-1,\hspace{1ex} 1]}]_{N-1}$  and
$[{\bm{S}}_{\overline{\bm{\gamma}}'}^{[N-1, \hspace{1ex} 1]}]_{N-1}$
and indices $i,j$ equal to the highest values $(i=N-2, j=N)$ for
${[\bm{S}}_{\overline{\bm{\gamma}}'}^{[N-2, \hspace{1ex} 2]}]_{i,j}$
  in order to
involve all $N$ particles in the motion.  Symmetry coordinates 
 with lower
values for $\xi$ or $i$ and $j$ would result in modified behavior 
involving fewer particles from that
analyzed in Section~\ref{sec:collective}.  This raises the question of what
behavior is actually dominant in an ensemble?  
A definitive answer to this question
is answered by obtaining the actual wave function for a particular
state of $N$ bosons or fermions with the correct permutation symmetry enforced
and looking at the dominant contributions from the degenerate
normal modes of the $[N-1,1]$ or $[N-2,2]$ sectors. 
Some light can be shed on this
question by noting that as $N$ increases a higher and higher percentage
of the degenerate symmetry coordinates in the $[N-1,1]$ and $[N-2,2]$
sectors will have evolved into the ``large $N$'' 
collective behavior as described above.
For example, for $N = 10$, over 80 \% of the degenerate
symmetry
coordinates in the $[N-1,1]$ sectors have evolved into motion
that is more appropriately described as single particle radial excitation or
single particle angular excitation behavior rather than an antisymmetric 
stretching
or bending motion.  
For $N=100$, over 90\% of these degenerate symmetry coordinates
have evolved into ``large $N$" collective behavior. 
Similarly in the $[N-2,2]$ sector, for $N=10$,  over 80\% of the
$N(N-3)/2$ degenerate symmetry coordinates have evolved
into  ``large $N$'' collective behavior with a single dominant
interparticle angle and significantly smaller responses from the remaining
interparticle angles.  For $N =100$, the percentage is over 90\% and
by $N=1000$, over 99\% of the degenerate symmetry coordinates
in this sector have ``large N'' collective behavior.

\subparagraph{Conclusions.}

This investigation into the evolution of collective behavior as a function of
$N$ suggests that this type of collective behavior defined by the normal
modes of the system smoothly and quickly evolves from the well-known 
vibrational motions of small $N$ systems that have been characterized as
symmetric breathing and bending, asymmetric stretching and bending and the
simultaneous opening and closing of interparticle angles, to the ``large $N$''
collective behavior that is more appropriately described as breathing,
center of mass, radial and angular particle-hole excitations and phonon.
Thus, the analysis of behavior for these five analytic expressions for the
normal mode solutions for a confined system of $N$ identical particles
yields consistent, physically intuitive behaviors that have been observed in
the laboratory.
The transition to ``large $N$'' collective behavior happens at very low values
of $N$ e.g. $N=10$, which is consistent with the good agreement obtained in 
previous few-body studies for thermodynamic 
quantities\cite{adhikari,hu6,hu7,hu8,blume1,blume2,levinsen,grining}.

What are the dynamics that can drive one of these collective behaviors  
to become the dominant 
behavior of a system
with competing behaviors, collective or not, suppressed? My analysis of
the $N$ dependence of the symmetry coordinates for a Hamiltonian that is
known to support collective behavior in the form of superfluidity 
at ultracold temperatures in the unitary regime has revealed two phenomena
that have the potential to support the creation and stabilization of 
collective behavior.
First the mixing of radial and angular behavior in the normal modes is seen to
limit to pure radial or pure angular behavior for very large $N$ resulting
in symmetry coordinates that are eigenfunctions of an approximate Hamiltonian
governing the physics of the unitary regime, thus acquiring some amount of 
stability if unperturbed. Second,
from Figs.~(\ref{fig:five})-(\ref{fig:seven}), 
one can see that for low values of $N$ the
five different frequencies start out closer in value, but as $N$ increases 
these five frequencies spread out creating large gaps between the values of
these five frequencies.  These gaps could provide the stability for collective
behavior if mechanisms to prevent the transfer of energy to
other modes exist (such as low temperatures) or can be 
constructed.

\section{Acknowledgments}
I would like to thank the National Science Foundation for financial support
under Grant No. PHY-1607544.

\appendix
\renewcommand{\theequation}{A\arabic{equation}}
\setcounter{equation}{0}

\section{The Symmetric Group and the Theory of Group Characters}\label{app:Char}
A group of transformations is a set of transformations which
satisfy the composition law $ab=c$ where $a$ and $b$ are any two
elements of the group and $c$ also belongs to the group. A group
also contains the identity element $I$ such that $aI=Ia=a$ and for
every element in the group its inverse is also to be found in the
group.

The symmetric group $S_N$ is the group of all permutations of $N$
objects and, as such, has $N!$ elements. A permutation may be
written as
\begin{equation}\label{eq:permutation}
\left(
\begin{array}{c@{\hspace{1.0ex}}c@{\hspace{1.0ex}}c@{\hspace{1.0ex}}c@{\hspace{1.0ex}}
c@{\hspace{1.0ex}}c@{\hspace{1.0ex}}c@{\hspace{1.0ex}}c}
1&2&3&4&5&6&7&8 \\
2&3&1&5&4&7&6&8 \end{array} \right) \end{equation}
This denotes the following transformation
\begin{center}
\begin{tabular}{|ccc|} \hline
\begin{tabular}{l}Object before \\ transformation \end{tabular} &
$\longrightarrow$ & \begin{tabular}{l} is transformed \\ to object
\end{tabular} \\ \hline
1 & $\longrightarrow$ & 2 \\
2 & $\longrightarrow$ & 3 \\
3 & $\longrightarrow$ & 1 \\
4 & $\longrightarrow$ & 5 \\
5 & $\longrightarrow$ & 4 \\
6 & $\longrightarrow$ & 7 \\
7 & $\longrightarrow$ & 6 \\
8 & $\longrightarrow$ & 8 \\ \hline \end{tabular}
\end{center}
A cycle is a particular kind of permutation where the object
labels are permuted into each other cyclically. For example the
cycle $(abc)$ means that object $a$ is transformed into object
$b$, object $b$ is transformed into object $c$ and object $c$ is
transformed into object $a$. The cycle $(abc)$ is termed a 3-cycle
since it cycles three objects. Like wise $(3479)$ is a 4-cycle and
$(5)$ is a 1-cycle, the latter transforms object five into itself.
All $N!$ permutations of the group $S_N$ may be decomposed into
cycles. For example, the permutation of Eq.~(\ref{eq:permutation})
may be decomposed into cycles as
\begin{equation}\label{eq:decomposition}
\left(
\begin{array}{c@{\hspace{1.0ex}}c@{\hspace{1.0ex}}c@{\hspace{1.0ex}}
c@{\hspace{1.0ex}}c@{\hspace{1.0ex}}c@{\hspace{1.0ex}}c@{\hspace{1.0ex}}c}
1&2&3&4&5&6&7&8 \\
2&3&1&5&4&7&6&8 \end{array} \right) = (123)(45)(67)(8)
\end{equation}
This consists of one 1-cycle, two 2-cycles and one 3-cycle. We can
denote the cycle structure of a permutation by the symbol
$(1^{\nu_1},2^{\nu_2},3^{\nu_3},\ldots,N^{\nu_N})$,\cite{hamermesh}
where the notation $j^{\nu_j}$ means a cycle of length $j$ and
$\nu_j$ equals the number of cycles of length $j$ in that
permutation. In the case of the permutation of
Eq.~(\ref{eq:decomposition}), its cycle structure is
$(1^1,2^2,3^1)$.

A matrix representation of a group is a set of nonsingular
matrices, including the unit matrix, which has the same
composition law as the elements of the group. The character of an
element of an matrix representation of a group is the trace of the
matrix. The character, as the trace of a matrix, is invariant
under a similarity transformation and thus elements of equivalent
representations have the same character. The set of all the
distinct characters of the elements of an irreducible
representation of the group uniquely specify the irreducible
representation up to an equivalence transformation. The characters
of irreducible representations of a group are termed {\em simple}
characters. All elements of a group which are related by a
similarity transformation are said to belong to the same class.
The character of the elements of a group belonging to a particular
class all have the same character. Thus there are as many distinct
characters for a group as there are classes.

For the group $S_N$ all elements with the same cycle structure
belong to the same class and so all elements of a matrix
representation of $S_N$ with the same cycle structure have the
same character.

A reducible matrix representation of a group may be bought to
block diagonal form by a similarity transformation, where the
individual blocks are irreducible matrix representations of the
same group with lower dimensionality. Thus the characters,
$\chi(R)$, of a reducible group are the sums of the characters of
the irreducible matrix representations into which it can be
decomposed, i.e.\
\begin{equation}\label{eq:chi_p}
\chi(R) = \sum_{p} \chi_p(R) \,,
\end{equation}
where $R$ denotes the element of the group, $p$ labels all of the
irreducible blocks into which the reducible matrix representation
of the group may be decomposed and $\chi_p(R)$ is the character of
the irreducible representation of the block labelled by $p$. Now
in a particular reducible representation a given irreducible
representation may be repeated along the diagonal $a_\alpha$ times
and so Eq.~(\ref{eq:chi_p}) may be rewritten as
\begin{equation}\label{eq:chi_a_alpha}
\chi(R) = \sum_{\alpha} a_\alpha \chi_\alpha(R) \,.
\end{equation}
The decomposition of $\chi(R)$ of Eq.~(\ref{eq:chi_a_alpha}) into
simple characters $\chi_\alpha(R)$ is unique, i.e.\ there is not
another decomposition of the form
\begin{equation}\label{eq:chi_b_alpha}
\chi(R) = \sum_{\alpha} b_\alpha \chi_\alpha(R) \,,
\end{equation}
where at least one of the $b_\alpha$ is different from the
corresponding $a_\alpha$. This follows from the fact that quite
generally
\begin{equation}\label{eq:num_rep}
a_\alpha = \frac{1}{h} \sum_R \chi^*_\alpha(R) \, \chi(R)
\end{equation}
and that the simple characters satisfy the orthogonality condition
\begin{equation}\label{eq:chi_orth}
\sum_R \chi^*_\alpha(R) \, \chi_\beta(R) = h \, \delta_{\alpha
\beta} \,,
\end{equation}
where $h$ is the number of elements in the group. For
Eqs.~(\ref{eq:chi_b_alpha}), (\ref{eq:num_rep}) and
(\ref{eq:chi_orth}) to be consistent we must have
\begin{equation}\label{eq:beqa}
b_\alpha = a_\alpha \,.
\end{equation}

The irreducible matrix representations of $S_N$ may be labelled by
a Young diagram ( = Young pattern = Young shape). A Young diagram
is a is a set of $N$ adjacent squares such that the row below a
given row of squares is equal to or shorter in length. The set of
all Young diagrams that can be formed from $N$ boxes of all
possible irreducible matrix representations of $S_N$.

A given Young diagram may be denoted by a partition. A partition,
$[\lambda_1, \hspace{1ex} \lambda_2, \hspace{1ex} \lambda_3,
\hspace{1ex} \ldots, \hspace{1ex} \lambda_N]$ is a series of $N$
numbers $\lambda_i$ such that $\lambda_1 \geq \lambda_2 \geq
\lambda_3 \geq \ldots \geq \lambda_N$ such that $\lambda_1 +
\lambda_2 + \lambda_3 + \cdots + \lambda_N = N$. The number
$\lambda_i$ is the number of boxes in row $i$ of the Young
diagram. Thus the set of all possible partitions of length $N$
labels all of the possible irreducible matrix representations of
$S_N$ and so the irreducible representation labels $\alpha$ and
$\beta$ in Eqs.~(\ref{eq:chi_a_alpha}), (\ref{eq:chi_b_alpha}),
(\ref{eq:num_rep}), (\ref{eq:chi_orth}) and (\ref{eq:beqa}) above,
for the group $S_N$ may be taken to run over the set of all
possible partitions.

A Young diagram can have up to $N$ rows. For an $N$-row
(one-column) Young diagram all of the $\lambda_i$s are non zero.
However only one Young diagram will have $N$ row; all of the rest
will have less than $N$ rows. Thus in all but one case the last
few $\lambda_i$s will be zero. It is standard practice to drop the
zeros and use a partition with less than $N$ numbers. Thus the
partition $[3,0,0]$ labelling an irreducible representation of
$S_3$ is usually abbreviated to just $[3]$.

\renewcommand{\theequation}{B\arabic{equation}}
\setcounter{equation}{0}

\section{Calculation of 
the ${\bm G}$ and ${\bm {FG}}$ matrix elements in the
symmetry-coordinate basis for the $[N]$ sector.}\label{app:sigmacalc} In this Appendix
we use the $W_{\bm{X}'}^{[N]}$ matrices ($\bm{X}'= \bar{\bm{r}}'$
or $\overline{\bm{\gamma}}'$) to calculate the $\bm{G}$
and $\bm{FG}$ matrix elements in the symmetry-coordinate basis,
$[\bm{\sigma_{[N]}^{\bm{G}}}]_{\bm{X}'_1,\,\bm{X}'_2}$ and
$[\bm{\sigma_{[N]}^{\bm{FG}}}]_{\bm{X}'_1,\,\bm{X}'_2}$\,, using:
\begin{equation}
[\bm{\sigma_{[N]}^Q}]_{\bm{X}'_1,\,\bm{X}'_2} =
(W_{\bm{X}'_1}^{[N]})_\xi \, {\bf Q}_{\bm{X}'_1 \bm{X}'_2} \,
[(W_{{\bm{X}'}_{2}}^{[N]})_\xi]^T \,.
\end{equation}
where $\bm{Q} = \bf G$ or $\bf FG$ and $\xi$ is a row label.
These elements
are used to obtain the mixing angles in Eq.~(\ref{eq:tanthetaalphapm}),
the normal mode frequencies in Eqs.~(\ref{eq:lambda12pm}) and 
(\ref{eq:abcalpha}), and the normalization coefficients in
Eqs.~(\ref{eq:calphapm}) and (\ref{eq:calpha2}).

\noindent Using Eq.~(\ref{eq:WNeqsqrtN1})
\begin{equation}
[W^{[N]}_{\bar{\bm{r}}'}]_i = \frac{1}{\sqrt{N}} \,
[{\bm{1}}_{\bar{\bm{r}}'}]_i
\end{equation}
and Eq.~(\ref{eq:WNgeqsqrt2ontnm11})
\begin{equation}
[W^{[N]}_{\overline{\bm{\gamma}}'}]_{ij} = \sqrt{\frac{2}{N(N-1)}}
\,\, [{\bm{1}}_{\overline{\bm{\gamma}}'}]_{ij}
\end{equation}
with Eq.(28) in Ref.~\cite{paperI}
\begin{equation}
\bm{G} = \left( \begin{array}{cc}
{\bf I}_N & \bm{0} \\
\bm{0} & \tilde{g}' {\bf I}_M + \tilde{h}' {\bf R}^T {\bf R}
\end{array} \right) \,,
\end{equation}
where $G$ is an $N(N+1)/2 \times N(N+1)/2$ matrix in the internal displacement
coordinates, 
${\bf I}_N$ is an $N \times N$ identity matrix, ${\bf I}_M$ is an $M \times M$ 
identity matrix with $M = N(N-1)/2$, 
${\bf R}$ is an $N \times M$
matrix such that ${R}_{i,jk}=\delta_{ij}+\delta_{ik}$, and $ \tilde{g}'$ and
$\tilde{h}'$ are defined in Ref.~\cite{paperI} in Eq.(29), we find:
\renewcommand{\jot}{1em}
\begin{eqnarray}
[\bm{\sigma_{[N]}^G}]_{\bar{\bm{r}}',\,\bar{\bm{r}}'} & = &
\sum_{i,j=1}^N [W_{\bar{\bm{r}}'}^{[N]}]_{i} \, [{\bm
G}_{\bar{\bm{r}}' \bar{\bm{r}}'}]_{ij}
\, [(W_{\bar{\bm{r}}'}^{[N]})^{T}]_{j} \,  \nonumber \\
& = & \frac{1}{N} \sum_{i,j=1}^N [{\bm{1}}_{\bar{\bm{r}}'}]_{i} \,
\delta_{ij} [{\bm{1}}_{\bar{\bm{r}}'}]_{j}  \nonumber \\
& = & \frac{1}{N} \sum_{i}^N 1  \nonumber \\
& = & 1 \,,
\end{eqnarray}
\renewcommand{\jot}{0em}
\renewcommand{\jot}{1em}
\begin{eqnarray}
[\bm{\sigma_{[N]}^G}]_{\bar{\bm{r}}',\,\overline{\bm{\gamma}}'}
 & = & \sum_{i=1}^N \sum_{k=2}^N \sum_{j=1}^{k-1}
[W_{\bar{\bm{r}}'}^{[N]}]_{i} \, [{\bm G}_{\bar{\bm{r}}'{\overline{\bm{\gamma}}'}}]_{i,\,jk}
\, [(W_{\overline{\bm{\gamma}}'}^{[N]})^{T}]_{jk} \,  \nonumber \\
& = & 0 \,,
\end{eqnarray}
\renewcommand{\jot}{0em}

\renewcommand{\jot}{1em}
\begin{eqnarray}
[\bm{\sigma_{[N]}^G}]_{\overline{\bm{\gamma}}',\, \bar{\bm{r}}'} 
& = & \sum_{j=2}^N \sum_{i=1}^{j-1} \sum_{k=1}^{N}
[W_{\overline{\bm{\gamma}}'}^{[N]}]_{ij} \, [{\bm
G}_{\overline{\bm{\gamma}}' \bar{\bm{r}}'}]_{ij,\,k}
\, [(W_{\bar{\bm{r}}'}^{[N]})^{T}]_{k} \,  \nonumber \\
& = & 0
\end{eqnarray}
\renewcommand{\jot}{0em}
and

\begin{widetext}
\begin{equation}
\begin{split}
[\bm{\sigma_{[N]}^G}]_{\overline{\bm{\gamma}}',\,\overline{\bm{\gamma}}'} 
& =  \sum_{j=2}^N \sum_{i=1}^{j-1}
\sum_{l=2}^{N} \sum_{k=1}^{l-1}
[W_{\overline{\bm{\gamma}}'}^{[N]}]_{ij} \, [{\bm
G}_{\overline{\bm{\gamma}}' \overline{\bm{\gamma}}'}]_{ij,\,kl}
\, [(W_{\overline{\bm{\gamma}}'}^{[N]})^{T}]_{kl} \,  \nonumber \\
& =   \frac{2}{N(N-1)} \sum_{j=2}^N \sum_{i= 1}^{j-1}
\sum_{l=2}^{N} \sum_{k=1}^{l-1}
    [{\bm{1}}_{\overline{\bm{\gamma}}'}]_{ij}
\bigl(\tilde{g'} (\delta_{ik} \delta_{jl} + \delta_{il}
\delta_{jk}) + \tilde{h'} (\delta_{ik} + \delta_{il} + \delta_{jk}
+ \delta_{jl}))
[{\bm{1}}_{\overline{\bm{\gamma}}'}]_{kl}   \nonumber \\
& =  \frac{2}{4N(N-1)} 
\renewcommand{\arraystretch}{1.5} 
\begin{array}[t]{l}
\Bigl(
{\displaystyle \sum_{i,j,k,l=1}^N \bigl[ \tilde{g'} (\delta_{ik}
\delta_{jl} + \delta_{il} \delta_{jk})
+ \tilde{h'} (\delta_{ik} + \delta_{il} + \delta_{jk} + \delta_{jl}) \bigr] }
\nonumber \\
{\displaystyle - \sum_{i,j,k = 1}^N \bigl[ \tilde{g'} (\delta_{ik}
\delta_{jk} + \delta_{ik} \delta_{jk})
+ \tilde{h'} (\delta_{ik} + \delta_{ik} + \delta_{jk} + \delta_{jk}) \bigr] } 
\nonumber \\
{\displaystyle - \sum_{i,k,l = 1}^N \bigl[ \tilde{g'} (\delta_{ik}
\delta_{il} + \delta_{il} \delta_{ik})
+ \tilde{h'} (\delta_{ik} + \delta_{il} + \delta_{ik} + \delta_{il}) \bigr] } 
\nonumber \\
{\displaystyle + \sum_{i,k = 1}^N \bigl[ \tilde{g'} (\delta_{ik}
\delta_{ik} + \delta_{ik} \delta_{ik}) + \tilde{h'} (\delta_{ik} +
\delta_{ik} + \delta_{ik} + \delta_{ik}) \bigr]}
\Bigr)
\end{array}
\renewcommand{\arraystretch}{1} \nonumber \\
& =   \frac{1}{2N(N-1)} \bigl[(2 \tilde{g'} N^2 - 4 \tilde{h'}
N^3) -2 (2 \tilde{g'} N + 4 \tilde{h'} N^2) + (2 \tilde{g'} N + 4
\tilde{h'} N) \bigr]  \nonumber \\
& =  \tilde{g'} + 2 \tilde{h'} (N-1) \,.
%
\end{split}
\end{equation}
\end{widetext}

Thus we obtain:
%
\begin{equation}
\bm{\sigma_{[N]}^{\bm{G}}} = \left(
\renewcommand{\arraystretch}{1.0}
\begin{array}{l@{\hspace{1.0em}}l} \protect[\bm{\sigma_{[N]}^{\bm{G}}}\protect]_{\bar{\bm{r}}',\,\bar{\bm{r}}'} =
1 & {\displaystyle
\protect[\bm{\sigma_{[N]}^{\bm{G}}}\protect]_{\bar{\bm{r}}',\,\overline{\bm{\gamma}}'} 
= 0} \\
{\displaystyle
\protect[\bm{\sigma_{[N]}^{\bm{G}}}\protect]_{\overline{\bm{\gamma}}',\,\bar{\bm{r}}'}
= 0} & {\displaystyle
\protect[\bm{\sigma_{[N]}^{\bm{G}}}\protect]_{\overline{\bm{\gamma}}',\,\overline{\bm{\gamma}}'}
= \left( \tilde{g}' + 2(N-1) \tilde{h}' \right)}
\end{array} \renewcommand{\arraystretch}{1} \right) \,.
\end{equation}
\noindent where $\bm{\sigma_{[N]}^{\bm{G}}}$ is a $2 \times 2$ matrix 
representation of $G$ for this $[N]$ sector, i.e.
in the basis of the two symmetry coordinates: ${\bm{S}}_{\bar{\bm{r}}'}^{[N]}$ and 
${\bm{S}}_{\overline{\bm{\gamma}}'}^{[N]}$.

Similarly letting ${\bf Q} = \bm{FG}$ and again 
using Eq.(28) in Ref.~\cite{paperI}:
\begin{equation}
\bm{FG}= \left(
\begin{array}{cc}
\tilde{a} {\bf I}_N + \tilde{b} {\bf J}_N & \tilde{e} {\bf R}
+ \tilde{f} {\bf J}_{NM} \\
\tilde{c} {\bf R}^T + \tilde{d} {\bf J}_{MN} & \tilde{g} {\bf I}_M
+ \tilde{h} {\bf R}^T {\bf R} + \tilde{\iota} {\bf J}_M
\end{array}\right) \,,
\end{equation}
where $\tilde{a}, \tilde{b}, \tilde{c}, \tilde{d}, \tilde{e}, \tilde{f},
\tilde{g}, \tilde{h}$, and $\tilde{iota}$ are defined in Ref.~\cite{paperI}
Eq. 29, we can derive:
\renewcommand{\jot}{1em}
\begin{eqnarray}
[\bm{\sigma_{[N]}^{\bm{FG}}}]_{\bar{\bm{r}}',\,\bar{\bm{r}}'} & =
& \sum_{i,j=1}^N [W_{\bar{\bm{r}}'}^{[N]}]_{i} \, [{\bm
FG}_{\bar{\bm{r}}' \bar{\bm{r}}'}]_{ij} \,
[(W_{\bar{\bm{r}}'}^{[N]})^{T}]_{j} \,  \nonumber \\
& = & \frac{1}{N} \sum_{i,j=1}^N [{\bm{1}}_{\bar{\bm{r}}'}]_{i} (
\tilde{a}
\delta_{ij} + \tilde{b} 1_{ij}) [{\bm{1}}_{\bar{\bm{r}}'}]_{j}  \nonumber \\
& = & \frac{1}{N} [ N \tilde{a} + \tilde{b} N^2 ]  \nonumber \\
& = & \tilde{a} + \tilde{b} N \,,
\end{eqnarray}
\renewcommand{\jot}{0em}

\begin{widetext}
\begin{equation}
\begin{split}
[{\bm{\sigma_{[N]}^{\bm{FG}}}}]_{\bar{\bm{r}}',\,\overline{\bm{\gamma}}'}
& =   \sum_{i=1}^N \sum_{k=2}^N \sum_{j=1}^{k-1}
[W_{\bar{\bm{r}}'}^{[N]}]_{i} \, [{\bm {FG}}_{\bar{\bm{r}}'
\overline{\bm{\gamma}}'}]_{i,\,jk}
\, [(W_{\overline{\bm{\gamma}}'}^{[N]})^{T}]_{jk}   
 =  \frac{1}{N} \sqrt{\frac{2}{N-1}} \sum_{i=1}^N \sum_{k=2}^N
\sum_{j=1}^{k-1} [{\bm{1}}_{\bar{\bm{r}}'}]_{i} \Bigl(\tilde{e}
(\delta_{ij} + \delta_{ik})
+ \tilde{f}  \Bigr) [{\bm{1}}_{\overline{\bm{\gamma}}'}]_{jk}  \nonumber \\
& =  \frac{1}{N \sqrt{2(N-1)}} \,\, \Bigl( \sum_{i,j,k=1}^N
\tilde{e} ( \delta_{ij} + \delta_{ik}) + \tilde{f} 
- \sum_{i,j}^N \tilde{e}
( \delta_{ij} + \delta_{ij}) + \tilde{f} \Bigr) \nonumber \\
& =   \frac{1}{N \sqrt{2 (N-1)}} \,\, ( 2 \tilde{e} N^2 +
\tilde{f} N^3
  - 2 \tilde{e} N - \tilde{f} N^2)  
 =  \sqrt{ 2 (N-1)} \, \left( \tilde{e} + \frac{N}{2} \tilde{f} \right) \,,\\
[\bm{\sigma_{[N]}^{\bm{FG}}}]_{\overline{\bm{\gamma}}',\,
\bar{\bm{r}}'} 
& =  \sum_{j=2}^N \sum_{i=1}^{j-1} \sum_{k=1}^{N}
[W_{\overline{\bm{\gamma}}'}^{[N]}]_{ij} \, [{\bm
FG}_{\overline{\bm{\gamma}}' \bar{\bm{r}}'}]_{ij,\,k} \,
[(W_{\bar{\bm{r}}'}^{[N]})^{T}]_{k} \,  
 =  \frac{1}{N} \sqrt{\frac{2}{N-1}} \sum_{j=2}^N
\sum_{i=1}^{j-1} \sum_{k=1}^N
[{\bm{1}}_{\overline{\bm{\gamma}}'}]_{ij} \Bigl[\tilde{c}
(\delta_{ki} + \delta_{kj})
+ \tilde{d}  \Bigr] [{\bm{1}}_{\bar{\bm{r}}'}]_{k}  \nonumber \\
& =  \frac{1}{N \sqrt{2(N-1)}} \,\, \Bigl( \sum_{i,j,k = 1}^N
\bigl[ \tilde{c} ( \delta_{ki} + \delta_{kj}) + \tilde{d} \bigr] -
\sum_{ik = 1}^N \bigl[ \tilde{c}
( \delta_{ki} + \delta_{ki}) + \tilde{d} \bigr] \Bigr)  \nonumber \\
& =  \frac{1}{N \sqrt{2 (N-1)}} \,\, ( 2 \tilde{c} N^2 +
\tilde{d} N^3
  - 2 \tilde{c} N - \tilde{d} N^2)  
 =  \sqrt{2(N-1)} \, \left(  \tilde{c} + \frac{N}{2} \,
\tilde{d} \right)
\end{split}
\end{equation}
\end{widetext}
and

\begin{widetext}
\begin{equation}
\begin{split}
[\bm{\sigma_{[N]}^{\bm{FG}}}]_{\overline{\bm{\gamma}}',\,
\overline{\bm{\gamma}}'} & =  \sum_{j=2}^N \sum_{i=1}^{j-1}
\sum_{l=2}^{N} \sum_{k=1}^{l-1}
[W_{\overline{\bm{\gamma}}'}^{[N]}]_{ij} \, [{\bm
FG}_{\overline{\bm{\gamma}}' \overline{\bm{\gamma}}'}]_{ij,\,kl}
\, [(W_{\overline{\bm{\gamma}}'}^{[N]})^{T}]_{kl} \,  \nonumber \\
& =  \frac{2}{N(N-1)} \Bigl( \sum_{j=2}^N \sum_{i= 1}^{j-1}
\sum_{l=2}^{N} \sum_{k=1}^{l-1}
[{\bm{1}}_{\overline{\bm{\gamma}}'}]_{ij} \bigl[ \tilde{g}
(\delta_{ik} \delta_{jl} + \delta_{il} \delta_{jk}) + \tilde{h}
(\delta_{ik} + \delta_{il} + \delta_{jk} + \delta_{jl})
+ \tilde{i} \bigr] [{\bm{1}}_{\overline{\bm{\gamma}}'}]_{kl} \Bigr)  \nonumber \\
& =  \frac{1}{2N(N-1)} 
\renewcommand{\arraystretch}{1.5} 
\begin{array}[t]{l}
\Bigl( 
{\displaystyle \sum_{i,j,k,l = 1}^N
 \bigl[ \tilde{g}
(\delta_{ik} \delta_{jl} + \delta_{il} \delta_{jk}) + \tilde{h}
(\delta_{ik} + \delta_{il} + \delta_{jk} + \delta_{jl})
+ \tilde{i} \bigr]  }  \nonumber \\
{\displaystyle - \sum_{i,j,k = 1}^N \bigl( \tilde{g} (\delta_{ik}
\delta_{jk} + \delta_{ik} \delta_{jk}) + \tilde{h} (\delta_{ik} +
\delta_{ik} + \delta_{jk} + \delta_{jk})
+ \tilde{i} \bigr) } \\
{\displaystyle - \sum_{i,k,l = 1}^N \bigl( \tilde{g} (\delta_{ik}
\delta_{il} + \delta_{il} \delta_{ik}) + \tilde{h} (\delta_{ik} +
\delta_{il} + \delta_{ik} + \delta_{il})
+ \tilde{i} \bigr)  } \nonumber \\
{\displaystyle + \sum_{i,k = 1}^N \bigl( \tilde{g} (\delta_{ik}
\delta_{ik} + \delta_{ik} \delta_{ik}) + \tilde{h} (\delta_{ik} +
\delta_{ik} + \delta_{ik} + \delta_{ik}) + \tilde{i} \bigr)}
\Bigr)
\end{array}
\renewcommand{\arraystretch}{1} \nonumber \\
& =  \frac{1}{2N(N-1)} \bigl(  2N(N-1) \tilde{g} + 4N( N^2-2N +
1) \tilde{h}
+ N^2 (N^2 - 2N + 1) \tilde{i} \bigr)  \nonumber \\
& =  \tilde{g} + 2 (N-1) \tilde{h} + \frac{N (N-1)}{2} \,\,
\tilde{i} \,.
\end{split}
\end{equation}
\end{widetext}
Thus we obtain the $2 \times 2$ matrix representation of $FG$ in the symmetry
basis of the $[N]$ sector.
%
\begin{widetext}
\begin{equation}
\bm{\sigma_{[N]}^{\bm{FG}}} = \left(
\renewcommand{\arraystretch}{1.5}
\begin{array}{l@{\hspace{1.5em}}l} \protect[\bm{\sigma_{[N]}^{\bm{FG}}}\protect]_{\bar{\bm{r}}',\,\bar{\bm{r}}'} =
(\tilde{a}+\tilde{b}N) & {\displaystyle
\protect[\bm{\sigma_{[N]}^{\bm{FG}}}\protect]_{\bar{\bm{r}}',\,
\overline{\bm{\gamma}}'} =
\sqrt{2(N-1)} \,\, \left( \tilde{e}+\frac{N}{2}\tilde{f} \right)} \\
{\displaystyle
\protect[\bm{\sigma_{[N]}^{\bm{FG}}}\protect]_{\overline{\bm{\gamma}}',\,\bar{\bm{r}}'}
= \sqrt{2(N-1)} \,\, \left( \tilde{c}+\frac{N}{2}\tilde{d}
\right)} & {\displaystyle
\protect[\bm{\sigma_{[N]}^{\bm{FG}}}\protect]_{\overline{\bm{\gamma}}',\,\overline{\bm{\gamma}}'}
= \left(
\tilde{g}+2(N-1)\tilde{h}+\frac{N(N-1)}{2}\,\,\tilde{\iota}
\right)}
\end{array} \renewcommand{\arraystretch}{1} \right) \,.
\end{equation}
\end{widetext}

\end{document}